\documentclass{gandb}
\usepackage{graphicx}
\usepackage{epsfig}
\usepackage{float}

\def\hal{{\textstyle \frac{1}{2}}}

\def\eff{\mathrm{eff}}
\def\mbx{\mathrm{max}}
\def\ex{\mathrm{ex}}
\def\sc{\mathrm{sc}}
\def\similar{(\mbox{similar y-terms})}
\def\simi{(\mbox{sim.~y-terms})}
\def\kv{\mbox{K-V} }
\def\<{\langle}
\def\>{\rangle}
\begin{document}
\evensidemargin 29pt
\fussy
\title{SELF-CONSISTENT DISTRIBUTIONS}
{THE PROBLEM OF SELF-CONSISTENT\\
PARTICLE PHASE SPACE DISTRIBUTIONS\\
FOR PERIODIC FOCUSING CHANNELS}
\author{J\"URGEN STRUCKMEIER and INGO HOFMANN}
\authorhead{J.~STRUCKMEIER AND I.~HOFMANN}
\address{\it Gesellschaft f\"ur Schwerionenforschung (GSI),
Postfach~11~05~52, 64220~Darmstadt, Germany}
\received{(Received 15 July 1992; in final form 26 August 1992)}
\abstract{Charged particle beams in equilibrium with their collective fields
are represented by self-consistent phase space distributions.
As the starting point, we assume the external focusing forces
to act {\em continuously\/} on the beam.
In the collisionless mean-field approximation, a broad class of
stationary self-consistent particle phase space distributions can then be constructed.
The method is reviewed how to determine the single-particle Hamiltonian of the
focusing system for a given phase space density function.
Subsequently, canonical transformations are used to relate this description
to a beam passing through a \emph{non-continuous} transport system.
The correspondence is exact for the linear K-V model and provides an
approximate matching prescription for distributions with nonlinear space-charge forces.
The transverse Hamiltonian generally changes with the focusing strength,
including in a rotationally symmetric periodic solenoid channel.
Our computer simulations, presented in the second part of the paper,
indicate very small rms-emittance changes for appropriately matched beams
in the continuous and periodic solenoid channels investigated.
The simulation for a periodic quadrupole channel yields a small persistent
increase of the rms-emittance, whose physical interpretation requires
control of numerical errors.}
\keywords{Self-consistent phase space distributions, space-charge,
focusing channels\newline~\newline
\hspace*{-22mm}Revised version of: Particle~Accelerators~{\bf 39}, 219--249 (1992)}
\maketitle
\section{INTRODUCTION}
The transport of intense charged particle beams is of
considerable interest for many applications.
That is why intensive theoretical and numerical studies
as well as experiments have been carried out on this subject.
A fundamental question concerns the existence of ``physically realistic''
self-consistent phase space distributions for \emph{periodic} beam transport systems.
In other words, is it possible to transport intense
charged particle beams -- at least in principle -- without any loss
of beam quality through periodic focusing channels?
The answer is especially important for applications where
the conservation of the beam quality is crucial.
This is the case for example in heavy ion fusion, where
beam losses cannot be tolerated.
We emphasize that ideal beam transport conditions are assumed, i.e.\
no field errors, fringe fields,
misalignment effects etc.\ are taken into account.
Furthermore, we require that no particle collisions occur and that
the paraxial ray approximation applies, i.e.\ the longitudinal
particle velocity is much larger than the transverse ones.
The basic moment equations do not require a prescribed particle phase space
distribution or a self-similar evolution of its profile.
In particular, self-similarity during variations of the beam cross section,
as assumed in Ref.~\cite{leecoo}, is not imposed on the subsequent physical evolution.
The scaling transformation used below specifies matched initial conditions,
but does not establish a self-similar exact evolution for a non-K-V beam.

For charged particle beams in transport devices with continuous external
focusing, we can easily set up the condition for a beam to be stationary.
Seen from the co-moving beam frame, the external forces that act on
the particles are constant, hence the total system is conservative.
For a time-independent self-consistent mean field, a particle distribution $f$ is stationary
if the phase space surfaces of constant energy $w$
are also surfaces of constant density\cite{lapo,gluck}, i.e.\ if $f=f(w)$.

For non-continuous focusing channels, an exact periodic solution for
unbunched beams is the so-called ``Kapchinskij-Vladimirskij''
\mbox{(K-V)} distribution\cite{kapvla}.
In continuous focusing, it is obtained if all particles are assumed to have
exactly the same transverse single-particle energy, with uniform density
on the corresponding phase space shell.
In a periodic channel, the shell is instead defined by the appropriate
Courant--Snyder invariants.
Here ``periodic'' means that the full distribution is reproduced at
corresponding positions in successive lattice periods,
$f(\mathbf z,s+S)=f(\mathbf z,s)$, with $\mathbf z=(x,y,x',y')$.
This requirement is stronger than periodicity of the second moments alone.
It has the advantage that the resulting equations
of motion turn out to be linear.
On the other hand this type of distribution is not ``physically realistic'',
since ``real'' beams always have some finite transverse energy spread.

Moreover, perturbations of this distribution can excite
instabilities\cite{hola,hofm} in parameter regimes where smoother
beam models show a more stable evolution.
The exact K-V distribution remains an equilibrium solution,
but its singular phase space structure can limit its usefulness
as a model of realistic high current beams.
Other models therefore need to be investigated.

After a short recall of the basic equations in section \ref{recall}, the
method of constructing stationary phase space distributions for
continuous beam transport systems is reviewed in section \ref{stationary}.
For unbunched beams a closed differential equation is
derived that allows us to directly determine the effective
potential, hence the system's Hamiltonian for any given phase space
density function $f = f(w)$.
We solve this differential equation for
the two simplest possible examples in section \ref{examp}.

In section \ref{catra}, a canonical transformation is suggested,
which performs in a formal way a transformation from a \emph{continuous}
focusing system into a \emph{periodic} system.
For the linear case, i.e.\ for a \kv particle phase space
distribution, this canonical transformation is just the inverse
of the transformation used earlier by Courant and Snyder\cite{cousny}
to transform from a periodic channel into a continuous.
It is also equivalent to the transformation theory developed
by Lewis\cite{lewis} and Leach\cite{leach} for the time-dependent
harmonic oscillator.

For non-K-V distributions, we distinguish a formal canonical transformation
from an exact correspondence between two self-consistent physical systems.
The nonlinear self-field prevents the direct extension of the K-V construction
to an arbitrary periodic focusing channel.
The transformation nevertheless provides an approximate matching prescription\cite{lund},
whose practical usefulness is examined numerically.
We also give the balance equation for the total transverse beam energy,
which is generally not conserved in an $s$-dependent focusing system.

In section \ref{initial}, we establish the preservation of a vanishing
initial rms-emittance derivative under a common transverse matching scale.
Finally in section \ref{final}, we re-derive an estimate of the maximum
emittance growth due to conversion of ``excess'' field energy for a
non-stationary initial distribution\cite{strekla}, assuming nearly constant rms radii.
Computer simulations shown in section \ref{simu} illustrate the equilibrium
construction, the matching prescription, and this energy-conversion estimate.
\section{RECAPITULATION OF THE BASIC EQUATIONS} \label{recall}
Before outlining the transformation theory, we recapitulate
some definitions and results needed in the subsequent sections.
We consider a charged particle beam that passes through a focusing system.
Gaussian electromagnetic units are used throughout.
The coordinates and slopes are measured relative to the beam centroid, so that the second moments below are centered moments.
For solenoid focusing, the transverse variables are understood in the rotating Larmor frame, in which the linear focusing equations have the uncoupled form used below.
The bunched-beam equations are understood as the nonrelativistic electrostatic model, $\gamma=1$.
For the transverse dynamics of unbunched paraxial beams, the displayed $\gamma$-dependent coefficients include the relativistic reduction of the space charge force.
The beam speed and the particle number are constant, and image-charge forces are neglected.
In general, regarded from the beam system,
the external forces are time-dependent.
As a consequence, the whole system is not conservative.
The total Hamiltonian $H$
of the system of $N$ particles of the charge $q$ interacting
through their Coulomb forces is given by
$$
H = m_0 c^2 \beta^2 \gamma \sum_{i=1}^{N}
\left( x_i'^2 + y_i'^2 + z_i'^2 \right) ~-~ L \; ,
$$
$L$ denoting the Lagrangian function
$$
L = T - \left( q V_{\ex} + \frac{1}{\gamma^2} W \right) \; ,
$$
with $T$ the kinetic energy of all particles, $q V_{\ex}$ the
potential energy resulting from the external focusing devices,
and $W$ the Coulomb interaction energy.
If the external focusing forces are linear,
the complete Hamiltonian is given by
\begin{eqnarray} \label{ham0}
H & = & \hal m_0 c^2 \beta^2 \gamma \sum_{i=1}^{N}
\left( x_i'^2 + y_i'^2 + z_i'^2 \right) \nonumber \\
& + & \hal m_0 c^2 \beta^2 \gamma \sum_{i=1}^{N}
\left( k_x^2(s) x_i^2 + k_y^2(s) y_i^2 + k_z^2(s) z_i^2 \right) +
\frac{q}{2\gamma^2} \sum_{i=1}^{N} V_{\sc}(x_i,y_i,z_i)\;~
\end{eqnarray}
with $V_{\sc}$ the space charge potential at the position of
the $i$-th particle
\begin{equation} \label{scpot}
V_{\sc}(x_i,y_i,z_i) = \sum_{j \neq i}
\frac{q}{\sqrt{(x_i-x_j)^2 + (y_i-y_j)^2 + (z_i-z_j)^2}} \; ,
\end{equation}
and $k_x^2(s)$, $k_y^2(s)$, $k_z^2(s)$ the focusing force functions in
$x$, $y$, $z$.

For convenience,
the time has been replaced by the distance $s = c \beta \cdot t$
along the focusing structure
$$
\dot{x}_i = c \beta \cdot x_i' \quad , \quad
\dot{p}_{x,i} = m_0 c^2 \beta^2 \gamma \cdot x_i'' \; ,
$$
and similarly in $y$ and $z$.

The equations of motion for particle $i$ result
from (\ref{ham0}) as
\begin{eqnarray} \label{bewe}
x_i'' + k_x^2(s) \, x_i - \frac{q}{m_0 c^2 \beta^2 \gamma^3}
E_x(x_i,y_i,z_i) & = & 0 \nonumber \\
y_i'' + k_y^2(s) \, y_i - \frac{q}{m_0 c^2 \beta^2 \gamma^3}
E_y(x_i,y_i,z_i) & = & 0           \\
z_i'' + k_z^2(s) \, z_i - \frac{q}{m_0 c^2 \beta^2 \gamma^3}
E_z(x_i,y_i,z_i) & = & 0 \; , \nonumber
\end{eqnarray}
with the electric field from space charge, given by
\begin{equation} \label{e_x}
E_x(x_i,y_i,z_i) = q \, \sum_{j \neq i} \frac{x_i - x_j}
{\bigl[ (x_i - x_j)^2 + (y_i - y_j)^2 + (z_i - z_j)^2 \bigr]^{3/2}} \; .
\end{equation}
With $N$ the total number of particles, the beam moments
can be defined as
\begin{eqnarray*}
\<x^2\>   & = & N^{-1} \cdot \sum_{i=1}^{N} x_i^2 \\
\<xx'\>   & = & N^{-1} \cdot \sum_{i=1}^{N} x_i x_i' \\
\<x'^2\>  & = & N^{-1} \cdot \sum_{i=1}^{N} x_i'^2 \\
\<xE_x\>  & = & N^{-1} \cdot \sum_{i=1}^{N} x_i E_x(x_i,y_i,z_i) \\
\<x'E_x\> & = & N^{-1} \cdot \sum_{i=1}^{N} x_i' E_x(x_i,y_i,z_i) \; .
\end{eqnarray*}
Using the equations of motion (\ref{bewe}), we
obtain for their derivatives
\begin{eqnarray} \label{deri}
\frac{d}{ds} \<x^2\> - \, 2\<xx'\> & = & 0 \nonumber \\
\frac{d}{ds} \<xx'\> - \<x'^2\> + \, k_x^2(s) \<x^2\> -
\frac{q}{m_0 c^2 \beta^2 \gamma^3} \<x E_x\> & = & 0 \\
\frac{d}{ds} \<x'^2\> + \, 2 k_x^2(s) \<xx'\> -
\frac{2q}{m_0 c^2 \beta^2 \gamma^3} \<x' E_x\> & = & 0 \nonumber \; .
\end{eqnarray}
With the following definition of the rms-emittance $\varepsilon_x(s)$
\begin{equation} \label{epsrms}
\varepsilon_x^2(s) ~=~ \<x^2\> \<x'^2\> - \<x x'\>^2 \; ,
\end{equation}
and (\ref{deri}) we can set up a differential equation
for the moment $\sqrt{\<x^2\>}$, which is proportional to the actual
beam width in $x$.
It is known as the ``rms envelope equation'' that has been
first derived by Sacherer\cite{sach}
\begin{equation} \label{kv1}
\frac{d^2}{ds^2} \sqrt{\<x^2\>} + k_x^2(s) \, \sqrt{\<x^2\>} -
\frac{q}{m_0 c^2 \beta^2 \gamma^3} \frac{\<xE_x\>}{\sqrt{\<x^2\>}} -
\frac{\varepsilon_x^2(s)}{\sqrt{\<x^2\>^3}} = 0 \; .
\end{equation}
A beam is called ``matched'' if the initial conditions of (\ref{kv1})
have been adjusted so that the rms-beam widths
have the same periodicities as the focusing functions $k_{x,y,z}^2(s)$.
Due to the fact that $\<xE_x\>$ and $\varepsilon_x^2$ are in general
functions of $s$, this equation is not closed.

The change of the rms-emittance is also readily
derived from (\ref{deri})
\begin{equation} \label{epsderi}
\frac{d}{ds} \varepsilon_x^2(s) = \frac{2q}{m_0 c^2 \beta^2 \gamma^3}
\; \bigl( \<x^2\> \<x'E_x\> - \<xx'\> \<xE_x\>\bigr) \; .
\end{equation}
The total interaction energy $W$ is half the sum over the
space charge potential contributions (\ref{scpot}) of all particles
\begin{eqnarray} \label{wgen}
W & = & \frac{q}{2} \, \sum_i V_{\sc}(x_i,y_i,z_i) \nonumber \\
& = & \frac{q^2}{2} \sum_i \sum_{j \neq i}
\frac{1}{\sqrt{(x_i-x_j)^2 + (y_i-y_j)^2 + (z_i-z_j)^2}} \; .
\end{eqnarray}
The beam moments $\<x' E_x\>$, $\<y' E_y\>$ and $\<z' E_z\>$ can be
correlated to the change of the interaction energy $W$, since
we are allowed to interchange the order of summation:
\begin{eqnarray} \label{dw}
\frac{dW}{ds} & = & -\frac{q^2}{2} \, \sum_i \sum_{j \neq i}
\frac{(x_i - x_j)(x_i' - x_j') + (y_i - y_j)
(y_i' - y_j') + (z_i - z_j)(z_i' - z_j')}
{\bigl[ (x_i - x_j)^2 + (y_i - y_j)^2 +
(z_i - z_j)^2 \bigr]^{3/2}} \nonumber \\
& = & -q \, \sum_i \bigl[
x'_i E_x(x_i,y_i,z_i) +
y'_i E_y(x_i,y_i,z_i) +
z'_i E_z(x_i,y_i,z_i) \bigr] \nonumber \\
& = & -N q \, \bigl( \<x' E_x\> + \<y' E_y\> + \<z' E_z\> \bigr) \; .
\end{eqnarray}
To define a uniform reference energy, we introduce a uniformly charged ellipsoid with the same total charge and the same second moments $\<x^2\>$, $\<y^2\>$, and $\<z^2\>$ as the actual beam.
Its axes are aligned with the coordinate axes, and averages over its particle density $n_u$ are denoted by $\<\cdot\>_u$.
The interior space charge field of this ellipsoid is linear in each coordinate.
Its coefficients are
$$
A_x^u(s)=\frac{\<xE_x^u\>_u}{\<x^2\>} \qquad , \qquad
A_y^u(s)=\frac{\<yE_y^u\>_u}{\<y^2\>} \qquad , \qquad
A_z^u(s)=\frac{\<zE_z^u\>_u}{\<z^2\>} \; .
$$
The corresponding interior potential is
\begin{equation} \label{vu1}
V_{\sc}^u(x,y,z;s)
= C_u(s)-\hal\left[A_x^u(s)x^2+A_y^u(s)y^2+A_z^u(s)z^2\right] \; .
\end{equation}
The factor $\hal$ is required by $E_x^u=-\partial V_{\sc}^u/\partial x$, and similarly in $y$ and $z$.
The constant $C_u(s)$ is determined by the same fixed potential reference as the actual self-potential.
It may be removed when only the force at a fixed $s$ is needed, but its derivative must be retained in an energy balance.
Moreover, (\ref{vu1}) is an interior formula and must not be extended to particles outside the reference ellipsoid as if it were the physical uniform-beam potential.

The field energy of the uniform reference beam is
$$
W_u(s)=\frac{q}{2}\int n_u(x,y,z;s)V_{\sc}^u(x,y,z;s)\,dx\,dy\,dz \; .
$$
This integral is evaluated with the reference density, rather than with the actual nonuniform particle distribution.
Following the changes of the reference ellipsoid's semiaxes, Eq.~(\ref{dw}) gives
\begin{equation} \label{dwu}
\frac{dW_u}{ds}=-Nq\left(
\frac{\<xx'\>}{\<x^2\>}\<xE_x^u\>_u+
\frac{\<yy'\>}{\<y^2\>}\<yE_y^u\>_u+
\frac{\<zz'\>}{\<z^2\>}\<zE_z^u\>_u\right) \; .
\end{equation}
In general, the field moments of this uniform reference beam differ from the corresponding moments of the actual beam.
Writing $(x_1,x_2,x_3)=(x,y,z)$ and $(\varepsilon_1,\varepsilon_2,\varepsilon_3)=(\varepsilon_x,\varepsilon_y,\varepsilon_z)$, Eqs.~(\ref{epsderi}), (\ref{dw}), and (\ref{dwu}) yield the exact identity within the stated Coulomb model
\begin{eqnarray} \label{sr2}
\sum_{\alpha=1}^{3}\frac{1}{\<x_\alpha^2\>}
\frac{d\varepsilon_\alpha^2}{ds}
&=&-\frac{2}{m_0c^2\beta^2\gamma^3N}
\frac{d}{ds}\bigl[W-W_u\bigr] \nonumber \\
&&\mbox{}+\frac{2q}{m_0c^2\beta^2\gamma^3}
\sum_{\alpha=1}^{3}\frac{\<x_\alpha x_\alpha'\>}{\<x_\alpha^2\>}
\left(\<x_\alpha E_\alpha^u\>_u-\<x_\alpha E_\alpha\>\right) \; .
\end{eqnarray}
The second term on the right must be included for a general three-dimensional distribution.
It contains the profile dependence that enters the generalized emittance equations discussed in Ref.~\cite{host}.

For the round unbunched beams considered below, the smooth-field approximation gives
$$
\<xE_x\>=\<yE_y\>=\frac{I}{2c\beta}\; ,
$$
independently of the radial charge density profile.
These moments agree with those of the equivalent uniform cylinder, so that the reference-moment correction vanishes and the longitudinal terms are absent.
We then obtain the exact transverse relation first derived for round beams by Wangler et al.~\cite{wang}
\begin{equation} \label{sr2_round}
\frac{1}{\<x^2\>}\frac{d\varepsilon_x^2}{ds}
+\frac{1}{\<y^2\>}\frac{d\varepsilon_y^2}{ds}
=-\frac{2}{m_0c^2\beta^2\gamma^3N}
\frac{d}{ds}\bigl[W-W_u\bigr] \; .
\end{equation}
For an unbunched beam, $N$ and the field energies refer to a fixed segment of length $L$, and the energy integral is replaced by $L\int dx\,dy$.
The logarithmic potentials of the actual and reference beams must use the same fixed reference length.
For example, with line charge $\lambda=I/(c\beta)=qN/L$, uniform radius $a_u=2\sqrt{\<x^2\>}$, and reference length $r_{\rm ref}$, the uniform energy is
$$
W_u=L\lambda^2\left(\ln\frac{r_{\rm ref}}{a_u}+\frac14\right) \; .
$$
It obeys $dW_u/ds=-L\lambda^2a_u'/a_u$, consistently with (\ref{dwu}) without its longitudinal term.
This size dependence would be lost if the potential were reset to zero at the beam center during the evolution.
\section{CONSTRUCTION OF STATIONARY PHASE SPACE DISTRIBUTIONS
IN CONTINUOUS FOCUSING CHANNELS} \label{stationary}
\subsection{Bunched Beams}
The bunched-beam equations are used within the nonrelativistic electrostatic scope specified in section~\ref{recall}, with $\gamma=1$.
The microscopic 6-dimensional phase space density is
$$
f_{\rm mic}(x,y,z,x',y',z',s) = \sum_{i=1}^{N} \delta(x - x_i) \cdot \:
\ldots \: \cdot \delta(z' - z'_i)
$$
and is normalized to the particle number $N$.
We describe the collective beam dynamics by its smooth one-particle mean-field approximation $f$, with the same normalization.
In the collisionless approximation, $f$ satisfies the Vlasov equation, which states that it is constant along the mean-field particle trajectories
\begin{equation} \label{liou}
\frac{d\/f}{ds} = 0 \; ,
\end{equation}
where the collective space charge forces are derived from a smooth self-consistent potential.
This approximation neglects collisional correlations, whereas Liouville's theorem for the full $N$-particle Hamiltonian system does not require the absence of interactions.

In order to set up the condition for a charged particle beam to be
stationary, we define $w$ as the scaled energy of
a particle moving inside the effective potential bucket
$V_{\eff}$ formed by the sum of the focusing potential and the
collective space charge potential
\begin{equation} \label{energy}
w(x,x',y,y',z,z') = \hal m_0 c^2 \beta^2 \gamma^3
\left( x'^2 + y'^2 + z'^2 \right) + q V_{\eff}(x,y,z) \; ,
\end{equation}
wherein
$$
qV_{\eff}(x,y,z) = \hal m_0 c^2 \beta^2 \gamma^3
\left( k_x^2 x^2 + k_y^2 y^2 + k_z^2 z^2 \right) +
q V_{\sc}(x,y,z) \; .
$$
Thus $w$ is $\gamma^2$ times the corresponding physical single-particle energy, with the assumed constant $\gamma$.
With the phase space energy function $w$, the Vlasov equation (\ref{liou}) can be rewritten as
$$
\frac{d\/f}{ds} \equiv \frac{\partial f}{\partial s}
+ \frac{\{f,w\}}{m_0 c^2 \beta^2 \gamma^3} = 0 \; ,
$$
where the bracket is defined in the coordinates and slopes by
$$
\{f,w\}=\sum_{u=x,y,z}
\left(\frac{\partial f}{\partial u}\frac{\partial w}{\partial u'}
-\frac{\partial f}{\partial u'}\frac{\partial w}{\partial u}\right).
$$
For constant external focusing and a time-independent self-consistent potential, the particle phase space distribution is stationary if its partial $s$-derivative vanishes
$$
\frac{\partial f}{\partial s} = 0 \; .
$$
Consequently, a vanishing Poisson bracket $\{f,w\}$ is a
sufficient condition for $f$ to be stationary.
This condition is always fulfilled if $f$ is a function of $w$
only, since a Poisson bracket $\{f(w),w\}$ vanishes identically
for arbitrary differentiable functions $f$ and $w$
$$
f = f(w) \; \Rightarrow \; f \; \mbox{is stationary} \; .
$$
In other words, a particle phase space distribution is stationary if
its phase space surfaces of constant energy
(``Isohamiltonians'') are also surfaces of constant density $f$.

In this context, the problem of ``self-consistency'' arises, since
due to the space charge potential $V_{\sc}$, the surfaces of constant
energy are, in turn, a function of the phase space distribution
$$
w = w(f) \; .
$$
With $q n(x,y,z)$ denoting
the macroscopic charge density function,
the space charge potential $V_{\sc}$ follows from Poisson's equation
\begin{equation} \label{poiss}
\Delta V_{\sc} = -4 \pi \, q \cdot n(x,y,z) \; .
\end{equation}
The particle density $n(x,y,z)$ in real space is obtained by
integrating the phase space distribution function over all velocities.
The self-consistent boundary of this integration is given by the
Isohamiltonian defined by the particle of the maximum allowable
energy $w_{\mbx}$
\begin{equation} \label{pro2}
n(x,y,z) = \int\limits_{-\zeta'}^{\zeta'}
\int\limits_{-\eta'}^{\eta'} \int\limits_{-\xi'}^{\xi'} \,
f(x,y,z,x',y',z') \; dx' \, dy' \, dz'
\end{equation}
with
\begin{eqnarray*}
\xi' ~\equiv  & x'_{\mbx}(x,y,z,y',z') & =~~\sqrt{\frac{2}{m_0 c^2\beta^2\gamma^3}
\bigl[ w_{\mbx} - q V_{\eff}(x,y,z) \bigr] - y'^2 - z'^2} \\
\eta' ~\equiv & y'_{\mbx}(x,y,z,z')~~~ & =~~\sqrt{\frac{2}{m_0 c^2\beta^2\gamma^3}
\bigl[ w_{\mbx} - q V_{\eff}(x,y,z) \bigr] - z'^2} \\
\zeta' ~\equiv &z'_{\mbx}(x,y,z)~~~~~~ & =~~\sqrt{\frac{2}{m_0 c^2\beta^2\gamma^3}
\bigl[ w_{\mbx} - q V_{\eff}(x,y,z) \bigr] } \; .
\end{eqnarray*}
Equation (\ref{pro2}) forms together with
Poisson's equation (\ref{poiss}) a closed set allowing the
determination of the self-consistent space charge potential
$V_{\sc}(x,y,z)$, since the type of distribution function
$f = f(w)$, the maximum energy $w_{\mbx}$
and the focusing constants $k_x$, $k_y$, $k_z$ are given.
\subsection{Unbunched Beams}
The procedure to construct a self-consistent phase space
distribution can be expressed in a simplified form if we consider a
4-dimensional transverse phase space distribution $f$ that describes
an unbunched beam in a rotationally symmetric focusing channel.
If $I$ denotes the longitudinal beam current, we normalize the transverse distribution $f$ to unity, so that
$$
\int f\,dx\,dy\,dx'\,dy'=1,
\qquad
\int n(x,y)\,dx\,dy=\frac{I}{qc\beta}.
$$
Using the abbreviations $r^2 = x^2 + y^2$ and $\rho^2 = x'^2 + y'^2$,
the energy function (\ref{energy}) takes on the form
\begin{eqnarray*}
w(r,\rho)
 & = & \hal m_0 c^2\beta^2\gamma^3\cdot \left( \rho^2 + k_0^2 \, r^2\right) + q V_{\sc}(r) \\
 & = & \hal m_0 c^2\beta^2\gamma^3\cdot\rho^2 + q V_{\eff}(r) \; .
\end{eqnarray*}
The projection (\ref{pro2}) can then be written as
\begin{equation} \label{pro3}
n(r) = \pi \frac{I}{q c \beta}
\int\limits_{0}^{\rho_{\mbx}^2(r)} f(r,\rho) \; d\/\left(\rho^2\right) \; ,
\end{equation}
with the self-consistent upper limit of integration
\begin{eqnarray}
\rho^2_{\mbx}(r) & = & \frac{2}{m_0 c^2 \beta^2 \gamma^3}
\bigl[ w_{\mbx} - q V_{\eff}(r) \bigr] \nonumber\\
& = & \frac{2q}{m_0 c^2 \beta^2 \gamma^3}
\bigl[ V_{\eff}(a) - V_{\eff}(r) \bigr] \label{rhomax}\; .
\end{eqnarray}
For the stationary distributions $f=f(w)$ considered here, we substitute $w$ as the new integration variable
\begin{eqnarray}
n(r) & = & \frac{2 \pi I}{q m_0 c^3 \beta^3 \gamma^3}
\int\limits_{q V_{\eff}(r)}^{q V_{\eff}(a)} \, f(w) \; dw \nonumber\\
& = & \frac{2 \pi I}{q m_0 c^3 \beta^3 \gamma^3}
\Bigl[ F\bigl( qV_{\eff}(a)\bigr) - F\bigl( qV_{\eff}(r)\bigr) \Bigr] \; ,
\label{wblcd}
\end{eqnarray}
if $F(w)$ denotes an integral function of
the given phase space distribution function $F^\prime(w)=f(w)$.
With this expression for the charge density, we can rewrite Poisson's
equation (\ref{poiss}) in cylindrical coordinates
to obtain an inhomogeneous differential equation for the
self-consistent effective potential $V_{\eff}(r)$:
\begin{equation} \label{intdif}
\frac{1}{r} \frac{d}{dr} \left( r \frac{d}{dr} qV_{\eff}(r) \right) -
4 \pi^2 K \cdot F\Bigl( qV_{\eff}(r)\Bigr)
= 2 m_0 c^2 \beta^2 \gamma^3 \cdot k_0^2 -
4 \pi^2 K \cdot F\Bigl( qV_{\eff}(a)\Bigr)\; .
\end{equation}
$K$ abbreviates the dimensionless ``generalized perveance'', defined as
\begin{equation} \label{perve}
K = \frac{2qI}{m_0 c^3 \beta^3 \gamma^3} \; .
\end{equation}
Starting with equation (\ref{intdif}), two examples how to construct
a self-consistent phase space distribution will be outlined
in the next section.
\section{EXAMPLES FOR SELF-CONSISTENT PARTICLE PHASE SPACE
DISTRIBUTIONS IN CONTINUOUS FOCUSING CHANNELS} \label{examp}
\subsection{``K-V'' Distribution} \label{kv}
For unbunched beams, the transverse phase space distribution that
defines all particles to have the same transverse total energy $w_{\mbx}$,
is commonly called \kv distribution.
Mathematically it is expressed using Dirac's $\delta$-distribution:
\begin{equation} \label{kvdef}
f(w)=\frac{m_0c^2\beta^2\gamma^3}{2\pi^2 a^2}\cdot\delta\bigl(w-w_{\mbx}\bigr)\; .
\end{equation}
The projection of the complete energy shell at $w_{\mbx}=qV_{\eff}(a)$ gives a uniform density inside $r<a$, thus (\ref{intdif}) becomes
\begin{equation} \label{kvgl1}
\frac{1}{r} \frac{d}{dr} \left( r \frac{d}{dr} qV_{\eff}(r) \right) =
2 m_0 c^2 \beta^2 \gamma^3 \cdot \left( k_0^2 - \frac{K}{a^2} \right) \; .
\end{equation}
The expression in parentheses
\begin{equation} \label{tune}
k^2 = k_0^2 - \frac{K}{a^2}
\end{equation}
shows that the zero current wave number $k_0$ is reduced
due to the space charge forces acting between the particles,
yielding the depressed tune of $k$.
For a nondegenerate finite-emittance equilibrium, we require $k^2>0$.
Nevertheless the right hand side is still a constant, so (\ref{kvgl1})
can easily be integrated:
$$
qV_{\eff}(r) = \hal m_0 c^2 \beta^2 \gamma^3 \cdot k^2r^2
\qquad (0\le r\le a) \; ,
$$
wherein the space charge potential $V_{\sc}$ has been normalized to vanish
on the beam axis:
$$
V_{\sc}(r) = -\frac{I}{c\beta} \cdot \frac{r^2}{a^2} \;\; .
$$
We observe that the self-consistent space charge potential is strictly
quadratic.
This leads to linear space charge forces and a homogeneous
space charge density \mbox{$qn_0 = I /(c\beta\pi a^2)$}.
Together with the linear focusing forces, the
equations of motion are completely linear.
The phase space energy function $w(r,\rho)$ for this case is given by:
$$
w(r,\rho) = \hal m_0 c^2 \beta^2 \gamma^3 \cdot \bigl( \rho^2 + k^2 r^2 \bigr)\;.
$$
The lines of constant energy (\ref{energy}) are thus
hyperellipsoids in phase space.
This reflects the fact that the net forces acting on all particles are linear.

The assumption that the particle beam coordinates are distributed
according to (\ref{kvdef}) is convenient because of the resulting
linear equations of motion.
On the other hand it has the disadvantage of being oversimplified from the
physical point of view.
The distribution of the transverse particle energies of a ``realistic'' beam
should be somewhat ``smeared'' and not as extremely peaked as suggested by (\ref{kvdef}).
This requirement is met by the ``water bag'' distribution
that will be discussed in the next section.
\subsection{``Water Bag'' Distribution} \label{water}
\begin{figure}[t]
\centering\epsfig{file=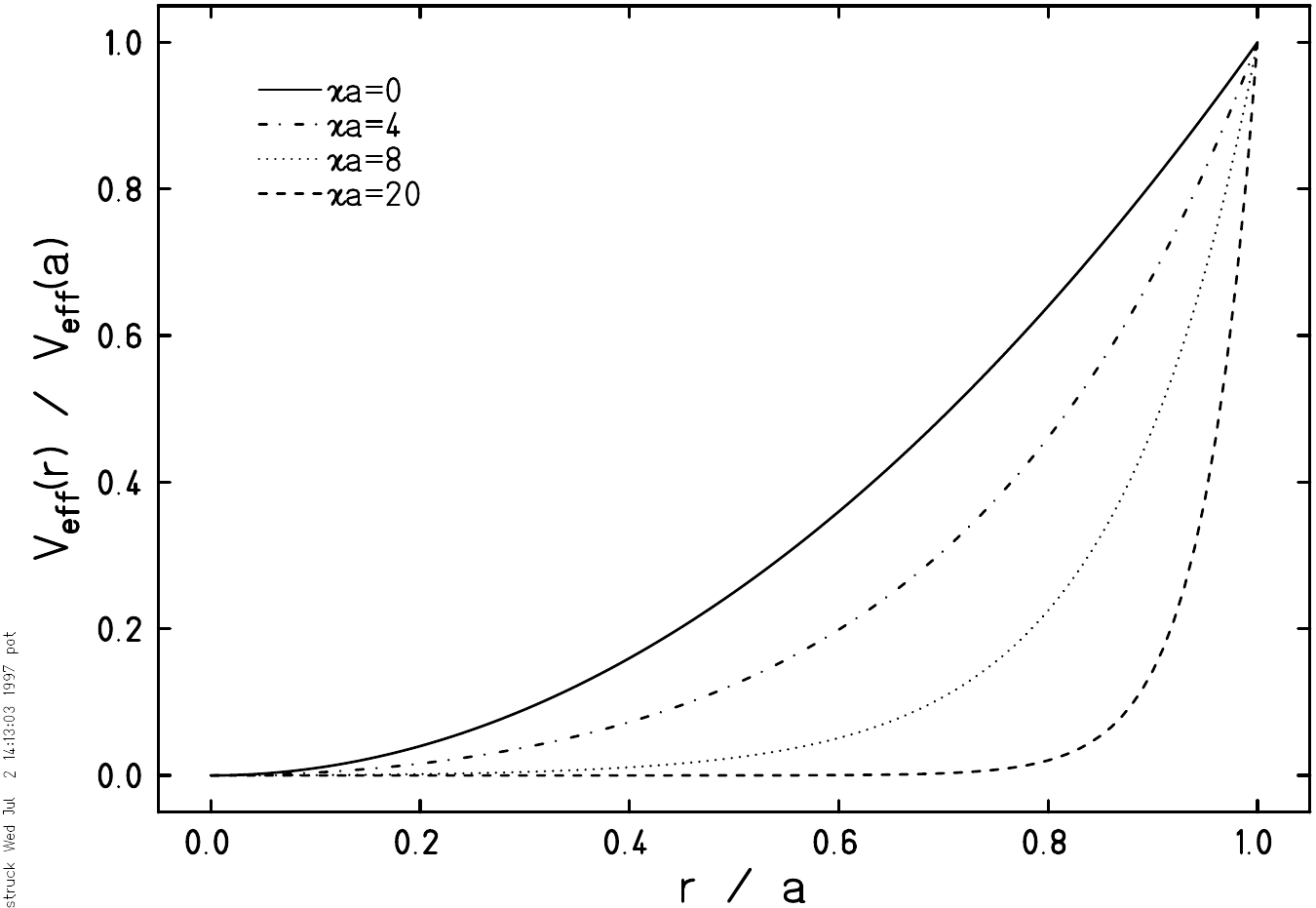,height=75mm}
\caption{Normalized effective potentials for stationary ``water bag''
distributions at different space charge parameters $\kappa a$.}
\label{effpotential}
\end{figure}

A particle distribution ansatz $f(w)$ that comes closer to reality with respect to exhibiting
the physical effect of ``Debye screening,'' while also allowing for an analytical treatment of
(\ref{intdif}) is the so-called ``water bag'' distribution\cite{lapo}.
For this case we assume a \emph{uniform} phase space density throughout the region between zero and a maximum energy $w_{\mbx}$.
Thus, equal phase space volume elements in that region carry equal probability, whereas no particles occur beyond $w_{\mbx}$.
The probability density of the particle energy also contains the density of available states and is therefore generally not uniform.
Mathematically this behavior can be expressed using the unit step function
$\Theta(x)$
\begin{equation} \label{wbdef}
f(w) = c_0\cdot \Theta \bigl( w_{\mbx} - w \bigr)
= \cases{c_0 &if $w \leq w_{\mbx}$ \cr ~ \cr 0 &if $w > w_{\mbx}$ \cr} \;\; ,
\end{equation}
with $c_0$ as normalization constant that will be determined later.
The integral function of (\ref{wbdef}) evaluates to $F(w)=c_0w$ up to an additive constant for
$w \leq w_{\mbx}$, thus equation (\ref{intdif}) becomes
\begin{equation} \label{wbgl1}
\frac{1}{r} \frac{d}{dr} \left( r \frac{d}{dr} qV_{\eff}(r) \right) =
2 m_0 c^2 \beta^2 \gamma^3 \cdot k_0^2 +
4 \pi^2 Kc_0\,q \cdot \Bigl[ V_{\eff}(r) - V_{\eff}(a) \Bigr] \;\; .
\end{equation}
In order to include the constants into a single expression,
we define temporarily the function $U(r)$ and replace the normalization constant $c_0$ by $\kappa^2$
\begin{equation} \label{bessel}
U(r) = V_{\eff}(r) - V_{\eff}(a) + \frac{2 m_0 c^2 \beta^2 \gamma^3 \cdot k_0^2}{q \cdot \kappa^2}
\quad {\rm with} \quad \kappa^2 = 4 \pi^2 K\,c_0 \;\; .
\end{equation}
In terms of $U(r)$, (\ref{wbgl1}) is converted into
a differential equation of the Bessel type
$$
\frac{1}{r} \frac{d}{dr} \left( r \frac{d}{dr} U(r) \right) -
\kappa^2 \cdot U(r) = 0 \;\; ,
$$
whose only physical solution for this case is the zeroth modified
Bessel function $I_0$
$$
U(r) = U(0)\cdot I_0(\kappa r) \, .
$$
With this solution and the definition $V_{\eff}(0) = 0$, we resolve (\ref{bessel}) for $V_{\eff}(r)$:
$$
V_{\eff}(r) = \frac{2 m_0 c^2 \beta^2 \gamma^3 \cdot k_0^2}{q \cdot \kappa^2} \cdot \frac{I_0(\kappa r) - 1}{I_0(\kappa a)}\,,
$$
hence
\begin{equation} \label{wbskpot}
V_{\eff}(r) = V_{\eff}(a) \cdot \frac{I_0(\kappa r) - 1}{I_0(\kappa a) - 1} \;,
\end{equation}
whereby the space charge potential has been adjusted to be zero on the beam axis.
The shapes of the normalized effective potentials $V_{\eff}(r)/V_{\eff}(a)$ for different
$\kappa a$ values are plotted in Fig.~\ref{effpotential}.

\begin{figure}
\centering\epsfig{file=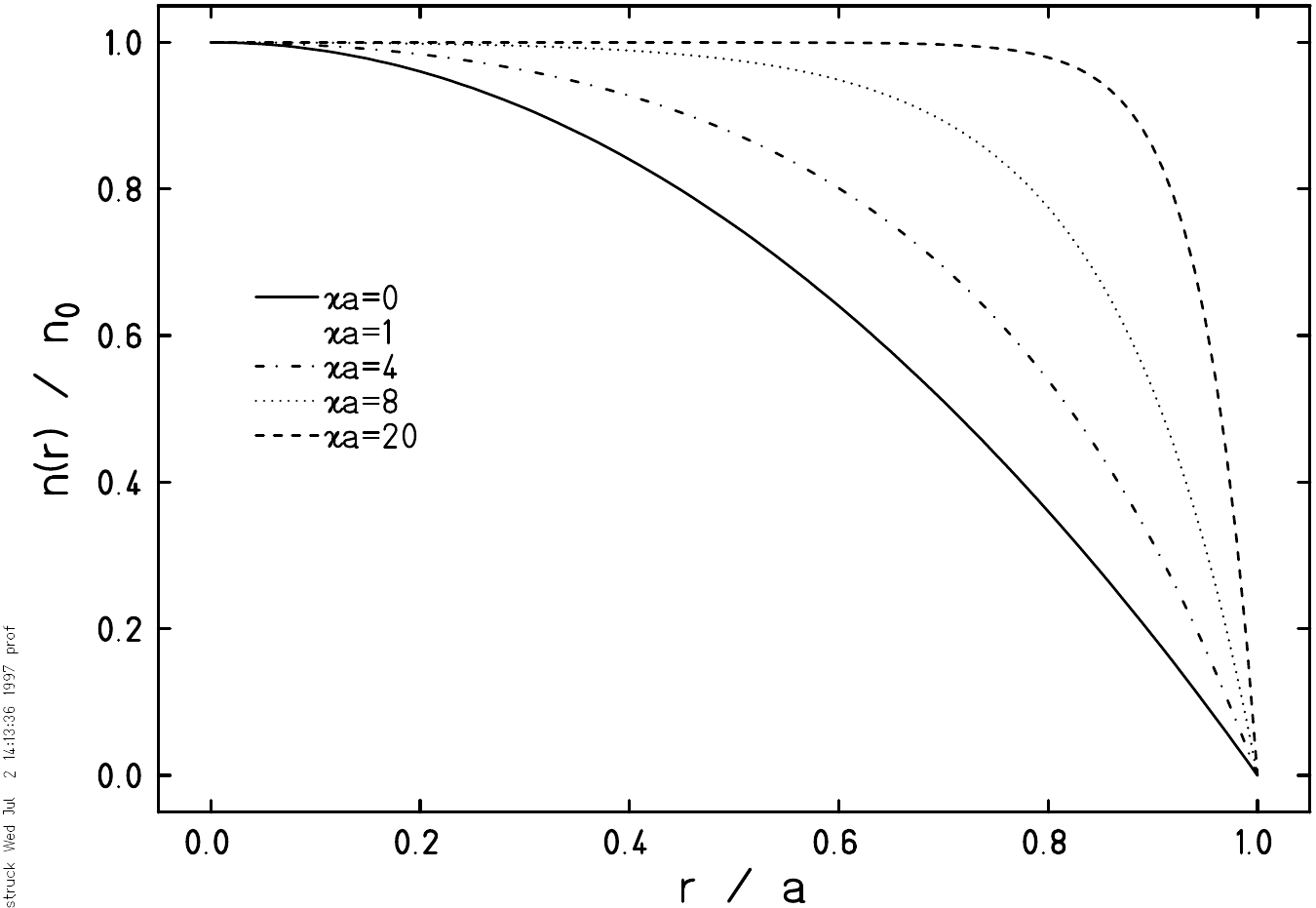,height=75mm}
\caption{Normalized radial charge densities for stationary ``water bag''
distributions at different space charge parameters $\kappa a$.}
\label{chargedensity}
\end{figure}
The density $n(r)$ now follows from Eq.~(\ref{wblcd}):
\begin{equation} \label{wbcd}
n(r) = n_0 \cdot \left( 1 - \frac{I_0(\kappa r)}{I_0(\kappa a)} \right)
\qquad (0\le r\le a)
\end{equation}
with the charge density constant $qn_0$:
$$
qn_0 = m_0 c^2 \beta^2 \gamma^3 \cdot \frac{k_0^2}{2\pi q} \; .
$$
The shapes of $n(r)/n_0$ for various values of $\kappa a$ are shown in Fig.~\ref{chargedensity}.
The space charge parameter $\kappa a$ can now be determined via normalization of the density function
$$
\frac{2\pi q c \beta}{I}\int_0^a \, n(r) \, r \, dr \stackrel{!}{=}1\,.
$$
By virtue of the identities
$$
\int_0^{\kappa a} uI_0(u)\,du=\kappa a\,I_1(\kappa a),\qquad
\frac{2I_1(\kappa a)}{\kappa a}=I_0(\kappa a)-I_2(\kappa a),
$$
the normalization of the density function~(\ref{wbcd}) leads to
the following implicit equation for the dimensionless space charge parameter $\kappa a$,
\begin{equation}\label{kappaa}
\frac{I_2(\kappa a)}{I_0(\kappa a)} = \frac{K}{(k_0a)^2} \;,
\end{equation}
which thus cannot be determined directly, but only via numerical solution.

A uniform charge density $qn_0$ produces a quadratic space charge
potential that is equal to the negative external focusing potential, thus
exactly cancels the latter.
The self-consistent maximum charge density, occurring on the beam axis,
is smaller by the factor $1 - 1/I_0(\kappa a)$ than the charge density $qn_0$
that fully screens out the external focusing potential at that point.
Due to the sharp rise of $I_0(\kappa a)$, this factor approaches unity in the strong-space-charge regime $\kappa a\gg1$.
In the weak-space-charge limit it instead tends to zero as $(\kappa a)^2/4$.

The effect that charged particles arrange themselves to screen out the external
potential is commonly designated in plasma physics as ``Debye shielding''.
We observe that --- in contrast to the \kv distribution ---
the ``water bag'' distribution exhibits this property.
Therefore, it is much better suited to simulate the behavior of real beams.

\begin{figure}
\centering\epsfig{file=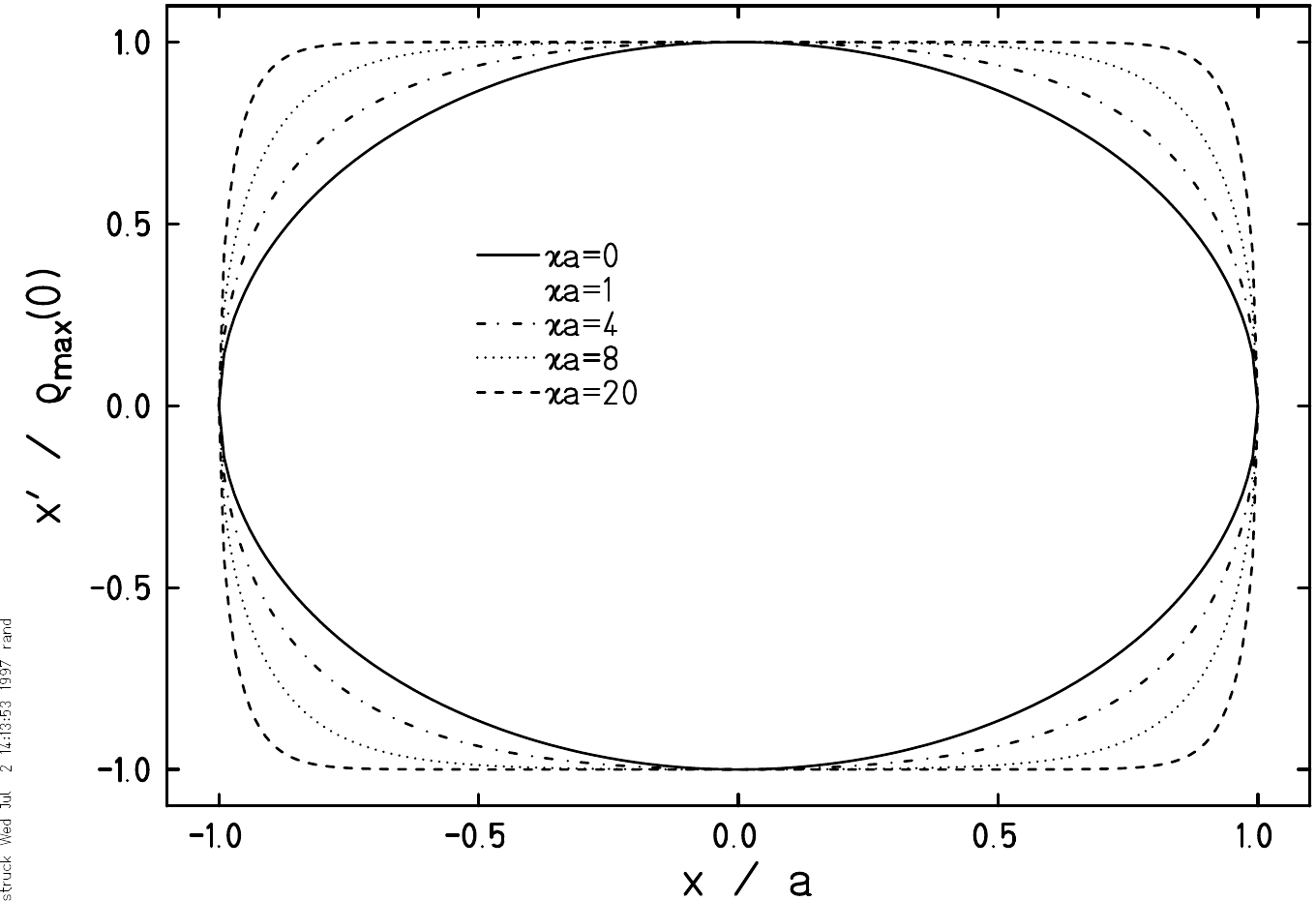,height=75mm}
\caption{Normalized phase space boundaries for stationary ``water bag''
distributions at different space charge parameters $\kappa a$.}
\label{phasespace}
\end{figure}
The surfaces of constant energy for the ``water bag'' distribution are
$$
w(r,\rho) = \hal m_0 c^2\beta^2\gamma^3\cdot\rho^2 + q V_{\eff}(a)
\cdot \frac{I_0(\kappa r) - 1}{I_0(\kappa a) - 1} \;.
$$
As self consistent phase space boundary follows
\begin{equation} \label{border}
\frac{\rho_{\mbx}^2(r)}{\rho_{\mbx}^2(0)} +
\frac{I_0(\kappa r) - 1}{I_0(\kappa a) - 1} = 1 \; .
\end{equation}
In the zero current limit, the effective potential turns out to be quadratic
$$
\lim_{\kappa \rightarrow 0} \, \frac{I_0(\kappa r) - 1}{I_0(\kappa a) - 1} =
\frac{r^2}{a^2} \; ,
$$
as expected for linear external focusing forces.
Thus Eq.~(\ref{border}) leads again to an ellipsoidal phase space symmetry.
In the strong-space-charge limit $\kappa a\to\infty$, the normalized potential term of (\ref{border}) exhibits a ``reflecting wall'' shape
$$
\lim_{\kappa \rightarrow \infty}\, \frac{I_0(\kappa r) - 1}{I_0(\kappa a) - 1} =
 \cases{0 &if $r < a$ \cr ~ \cr 1 &if $r = a$ \cr} \;\; .
$$
The applied focusing is almost canceled in the beam interior, and the particles turn within a narrow edge region.
The normalized phase space boundary therefore approaches a rectangle, as shown in Fig.~\ref{phasespace}.
This limit concerns strong relative space charge, not an infinite spatial density, since $n(r)\le n_0$ remains finite at fixed focusing strength.
\subsection{Generation of a ``Water Bag'' Phase Space Distribution for Simulations}
In order to compare the beam dynamics of different particle phase space distributions,
we draw upon the concept of ``equivalent beams'' introduced by Lapostolle\cite{lapo} and Sacherer\cite{sach}.
For our simulations, we therefore must generate distributions of given RMS parameters.
The beam radius then follows for the respective distribution as a function of the focusing strength $k_0$ and the scaled beam current $K$.
Before generating the phase space distribution, we must calculate the ratios of the second moments to the squared marginal beam sizes, i.e., the ratios
$$
\frac{\<x^2\>}{a^2},\qquad \frac{\<{x^\prime}^2\>}{\rho_{\max}^2(0)}.
$$
For an unbunched and transversely round beam of radius $a$, the second moment $\<x^2\>$ is defined as
$$
\<x^2\>=\frac{qc\beta}{I}\int\!\!\!\int x^2 n(x,y)\, dxdy
=\frac{\pi qc\beta}{I}\int_0^a r^3n(r)\,dr.
$$
This means for the density profile~(\ref{wbcd}) of the ``water bag'' distribution
\begin{eqnarray}
\frac{\<x^2\>}{a^2}&=&\frac{1}{{(\kappa a)}^4}\int_0^{\kappa a}\frac{I_0(\kappa a) - I_0(u)}{I_2(\kappa a)}u^3\,du\nonumber\\
&=&\frac{1}{2}+\frac{2}{{(\kappa a)}^2}-\frac{I_0(\kappa a)}{4I_2(\kappa a)}
=\cases{\to \frac{1}{6}\quad\mathrm{for}\quad\kappa\to0\cr ~\cr \to\frac{1}{4}\quad\mathrm{for}\quad\kappa\to\infty}
\label{rmsbs}
\end{eqnarray}
For a given RMS emittance $\varepsilon$, scaled beam current $K$, and zero current tune $k_0$, the stationary (matched) RMS beam size $\sqrt{\<x^2\>}$ is obtained from the RMS envelope equation\cite{sach}
\begin{equation}\label{rmsenvelope}
k_0^2\,\<x^2\> - \frac{K}{4} - \frac{\varepsilon^2}{\<x^2\>}=0.
\end{equation}
Combining now Eqs.~(\ref{kappaa}), (\ref{rmsbs}), and~(\ref{rmsenvelope}) yields the following implicit equation
for the dimensionless space charge parameter $\kappa a$ as a function of the given quantities $k_0, K, \varepsilon$:
\begin{equation}\label{kappaeps}
\frac{\frac{I_2(\kappa a)}{I_0(\kappa a)}}{1+\frac{4}{{(\kappa a)}^2}-\frac{I_0(\kappa a)}{2I_2(\kappa a)}}
=\frac{4}{1+\sqrt{1+u^2}},\qquad u=\frac{8k_0\,\varepsilon}{K}.
\end{equation}
For $K>0$, $k_0>0$, and $\varepsilon>0$, this equation can be solved numerically for $\kappa a$ via the \emph{regula falsi} method.
With the known space charge parameter $\kappa a$, the \emph{spatial} extension $a$ of the populated phase space can be calculated:
$$
a=\frac{\sqrt{K}}{k_0}\sqrt{\frac{I_0(\kappa a)}{I_2(\kappa a)}}.
$$
The corresponding ratio in velocity space must be calculated separately as it does not coincide with the spatial ratio in the case of non-zero beam current.
In general, the self-consistent upper limit of the normalized transverse velocity $\rho_{\max}(r)$ is given by Eq.~(\ref{rhomax}), hence in particular for the
self-consistent ``water bag'' distribution:
$$
\rho^2_{\max}(r)=4K\,\frac{I_0(\kappa a)-I_0(\kappa r)}{{(\kappa a)}^2 I_2(\kappa a)}.
$$
\begin{figure}[ht]
\centering\epsfig{file=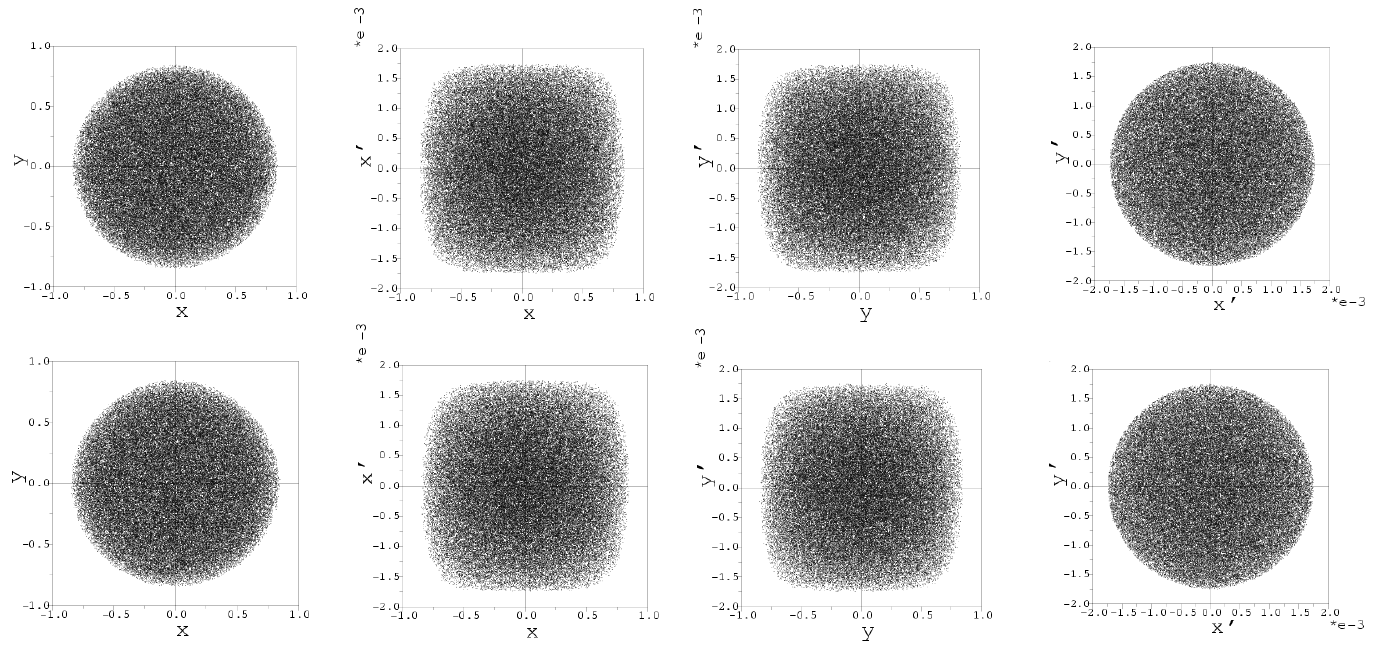,height=60mm,angle=0}
\caption{Self-consistent ``water bag'' distribution in a continuous focusing channel at $\sigma_0=60^\circ, \sigma=15^\circ$.
\underline{Upper row:} initial state, \underline{lower row:} after $15^\circ$ phase advance.}
\label{scwb-trans}
\end{figure}
The second moment $\<x^{\prime 2}\>$ then follows as
\begin{eqnarray*}
\<x^{\prime 2}\>&=&\frac{\frac{1}{2}\int\limits_{r=0}^a\int\limits_{\rho^2=0}^{\rho_{\max}^2(r)}\rho^2\,d(\rho^2)\,r\,dr}
{\int\limits_{r=0}^a\int\limits_{\rho^2=0}^{\rho_{\max}^2(r)}d(\rho^2)\,r\,dr}
=\frac{K}{{(\kappa a)}^2 I_2(\kappa a)}\frac{\int\limits_{u=0}^{\kappa a}{\left[I_0(\kappa a)-I_0(u)\right]}^2 u\,du}
{\int\limits_{u=0}^{\kappa a}{\left[I_0(\kappa a)-I_0(u)\right]} u\,du}\\
&=&\frac{2K}{{(\kappa a)}^2 I_2(\kappa a)}\left[I_0(\kappa a)-\frac{I_1^2(\kappa a)}{2I_2(\kappa a)}\right],
\end{eqnarray*}
and hence the ratio
$$
\frac{\<x^{\prime 2}\>}{\rho^2_{\max}(0)}=\frac{1}{2}\frac{1}{I_0(\kappa a)-1}\left[I_0(\kappa a)-\frac{I_1^2(\kappa a)}{2I_2(\kappa a)}\right]
=\cases{\to \frac{1}{6}\quad\mathrm{for}\quad\kappa\to0\cr ~\cr \to\frac{1}{4}\quad\mathrm{for}\quad\kappa\to\infty}
$$
We are now ready to generate a self-consistent ``water bag'' distribution
\begin{enumerate}
 \item calculate $\kappa a$ according to the implicit equation (\ref{kappaeps}) for the given parameter ratio $K/(k_0\,\varepsilon)$,
 \item generate random $4$-tuples $(\hat{x},\hat{y},\hat{x}^\prime,\hat{y}^\prime)$ with independent components uniformly distributed in $(-1,1)$, and retain only those that satisfy
 $$
 \hat{x}^{\prime 2}+\hat{y}^{\prime 2}+\frac{I_0(\kappa a\cdot\hat{r})-1}{I_0(\kappa a)-1}\le1,\qquad \hat{r}^2=\hat{x}^2+\hat{y}^2,
 $$
 \item calculate the unnormalized particle coordinates
 $$
 x = a\,\hat{x},\qquad y = a\,\hat{y},\qquad x^\prime=\rho_{\max}(0)\,\hat{x}^\prime,\qquad y^\prime=\rho_{\max}(0)\,\hat{y}^\prime,
 $$
 with
 $$
 a=\frac{\sqrt{K}}{k_0}\sqrt{\frac{I_0(\kappa a)}{I_2(\kappa a)}},\qquad\rho_{\max}(0)=\frac{2\sqrt{K}}{\kappa a}\sqrt{\frac{I_0(\kappa a)-1}{I_2(\kappa a)}}.
 $$
 \end{enumerate}
For $K=0$, the limiting distribution is generated by retaining tuples with $\hat{x}^2+\hat{y}^2+\hat{x}^{\prime 2}+\hat{y}^{\prime 2}\le1$ and using
$$
a=\sqrt{\frac{6\varepsilon}{k_0}},\qquad
\rho_{\max}(0)=\sqrt{6k_0\varepsilon}.
$$
A numerical verification of the self-consistency of ``water bag'' distribution generated this way is plotted in Fig.~\ref{scwb-trans}.

\subsection{``Semi-Gaussian'' distribution}
\begin{figure}[ht]
\hspace*{-5mm}\epsfig{file=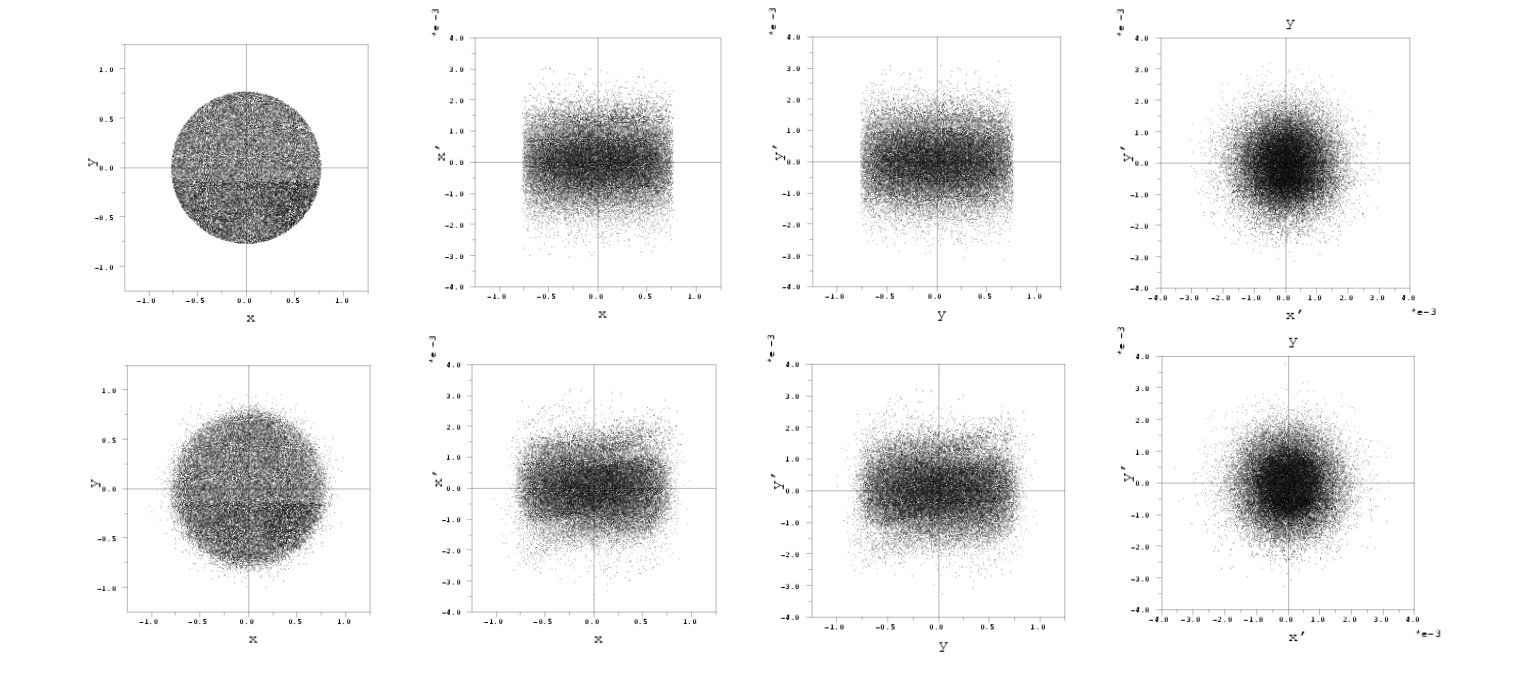,height=60mm,angle=0}
\caption{Semi-Gaussian distribution in a continuous focusing channel at $\sigma_0=60^\circ, \sigma=15^\circ$.
\underline{Upper row:} initial state, \underline{lower row:} after $15^\circ$ phase advance.}
\label{sg-trans}
\end{figure}

In the \emph{strong-space-charge regime}, the so-called ``semi-Gaussian'' distribution approximates the flat interior density and Gaussian velocity dependence of a self-consistent thermal equilibrium.
At finite temperature, its sharp spatial boundary makes it an approximate initial distribution rather than an exact stationary solution.
It has the following properties:
\begin{itemize}
 \item homogeneous density in real space with sharp boundary, hence $\sqrt{\<x^2\>/a^2}=0.5$,
 \item an isotropic Gaussian density in velocity space, truncated radially at $\rho_c=4\sigma_v$, where $\sigma_v$ denotes the RMS width of one velocity component before truncation, hence
 $$
 \frac{\sqrt{\<x^{\prime 2}\>}}{\rho_c}
 =\frac{1}{4}\sqrt{\frac{1-9e^{-8}}{1-e^{-8}}}
 \simeq0.2496642,
 $$
 \item rectangular boundaries in the $x,x^\prime$- and $y,y^\prime$-phase space projections.
\end{itemize}
A numerical simulation of the evolution of a ``semi-Gaussian'' distribution corresponding to the case of Fig.~\ref{scwb-trans}
is plotted in Fig.~\ref{sg-trans}.

\begin{figure}[ht]
\centering\epsfig{file=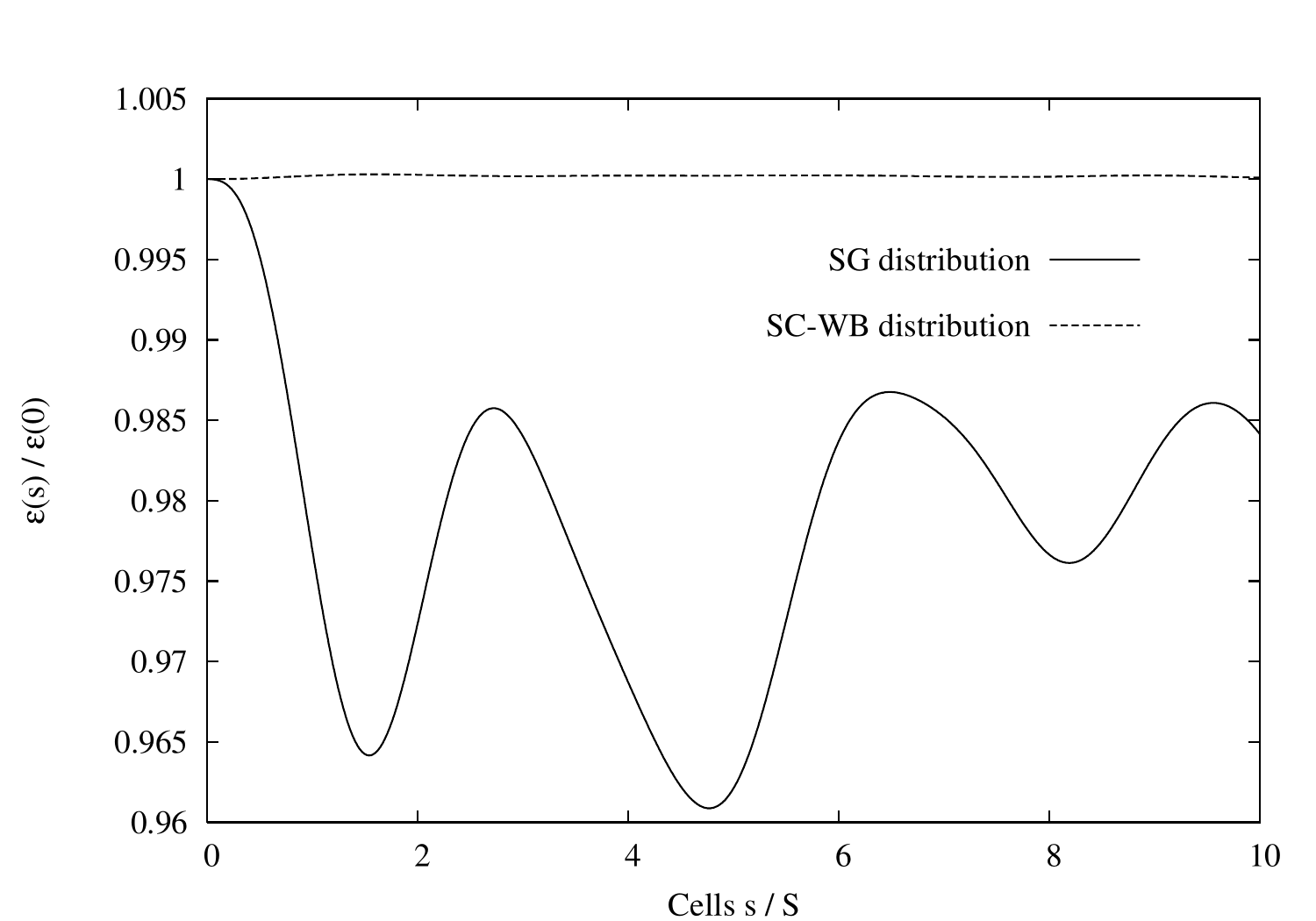,height=125mm,angle=-90}
\caption{RMS emittance growth factors versus $s$ for self-consistent ``water bag'' and semi-Gaussian distributions in a continuous focusing channel at $\sigma_0=60^\circ, \sigma=15^\circ$.}
\label{sg-emi}
\end{figure}
We observe that the initially exact rectangular shapes of the $x,x^\prime$- and $y,y^\prime$-phase space projections
evolve to rounded rectangles, similar to those of the self-consistent ``water bag'' distribution.
The initially sharp fringe of the $x,y$-projection gets blurred and thus exhibits the ``Debye shielding'' effect,
while the $x^\prime,y^\prime$-projection remains almost unchanged.

The evolutions of the RMS emittances of the ``semi-Gaussian'' (SG) distribution compared to the self-consistent ``water bag'' (SC-WB) distribution is plotted in Fig.~\ref{sg-emi}.
As expected, the RMS emittance of the SC-WB distribution stays constant.
In contrast, the RMS emittance of the SG distribution slightly shrinks due to the de-homogenization of the charge distribution in real space.
This effect --- investigated in detail in Sect.~\ref{final} --- is associated with an \emph{increase} of the associated electric field energy,
which is in turn associated with reduced transverse velocities and hence a reduction of the transverse emittance.
\section{CANONICAL TRANSFORMATION OF A CONTINUOUS CHANNEL INTO
A PERIODIC CHANNEL FOR UNBUNCHED BEAMS} \label{catra}
\subsection{Preliminaries}
As we have seen in the previous section, a self-consistent phase space distribution $f=f(w)$ is stationary in a continuous focusing channel when the effective potential is independent of $s$.
Its Isohamiltonians are then also surfaces of constant phase space density.

Realistic devices for charged particle optics consist of discrete lenses.
In order to construct suitable initial distributions for such devices, we shall correlate the particle coordinates in a continuous focusing channel with those in a system whose focusing forces depend on $s$.
For a \kv distribution the construction gives an exact correspondence between the single-particle equations of motion.
For a general distribution it provides a matching prescription whose dynamical accuracy must be assessed separately.
For the sake of transparency, the canonical mapping will be carried out in two steps.
The quantities obtained after the first step will be marked by a tilde.
In the same way, over-lined quantities denote those resulting from the second step.
It is understood that all quantities denoted with the index $x$ are defined likewise for the $y$-coordinate.

For unbunched beams, the charge density is independent of $z$.
If in addition no longitudinal focusing forces are acting, the Hamiltonian (\ref{ham0}) reduces to its transverse part $H_{\perp}$.
The corresponding two-dimensional interaction kernel is logarithmic.
For a numerical representation by $N_{\rm s}$ equal-current line elements, each carrying the current $I/N_{\rm s}$, the space charge potential is represented by
\begin{equation} \label{scpot_l}
V_{\sc}(x_i,y_i) \simeq -\frac{I}{N_{\rm s}c\beta}
\sum_{j\ne i}^{N_{\rm s}}
\ln\frac{(x_i-x_j)^2+(y_i-y_j)^2}{r_{\rm ref}^{\,2}} \; .
\end{equation}
Here $r_{\rm ref}$ is a fixed reference length, and $N_{\rm s}$ denotes the sampling number rather than the physical particle number $N$.
The smooth collective potential is obtained in the continuum limit of this representation.
Image forces are neglected in (\ref{scpot_l}).
For a centered, rotationally symmetric smooth beam inside a circular pipe, the choice $r_{\rm ref}=r_{max}$ makes the continuum potential vanish at the pipe wall.
For a general charge distribution, a conducting boundary requires its own Green function and cannot be imposed by a source-dependent normalization of the free-space logarithm.
The electric space charge field is represented correspondingly by
\begin{equation} \label{e_x_l}
E_x(x_i,y_i) \simeq \frac{2I}{N_{\rm s}c\beta}
\sum_{j\ne i}^{N_{\rm s}}
\frac{x_i-x_j}{(x_i-x_j)^2+(y_i-y_j)^2} \; .
\end{equation}

The canonical transformations below concern the single-particle characteristics in a prescribed smooth, self-consistent mean field.
Their generating Hamiltonian contains the full single-particle space charge potential, whereas the collective energy $H_{\perp}$ in (\ref{ham0}) contains one half of the summed potential contributions to avoid double counting.
To keep this distinction explicit, we write $h_{\perp,i}$ for the single-particle Hamiltonian and $\mathcal H_{\perp}=\sum_i h_{\perp,i}$ for the generator of the independent characteristics with the mean field held fixed when taking canonical derivatives.
The quantity $\mathcal H_{\perp}$ is not the collective transverse energy $H_{\perp}$.
Coordinate-independent potential terms can be omitted in the characteristic equations.
They must be retained with a common fixed reference when comparing collective energies.

For a stationary beam passing through a continuous focusing channel, $k_x^2$ and $\<x^2\>$ are constant, thus $\<xx'\>=0$.
Hence the second equation of (\ref{deri}) reduces to
\begin{equation} \label{kv2}
k_x^2 ~=~ \kappa_x^2+
\frac{q}{m_0c^2\beta^2\gamma^3}\,
\frac{\<xE_x\>}{\<x^2\>} \; ,
\end{equation}
with $\kappa_x^2=\<x'^2\>/\<x^2\>$.
Define the auxiliary quadratic potential and the remaining part of the continuous-channel self-potential by
\begin{eqnarray*}
V_{\sc}^{\rm quad}(x,y)
&=&-\hal\frac{\<xE_x\>}{\<x^2\>}x^2
 -\hal\frac{\<yE_y\>}{\<y^2\>}y^2,
\\
\Delta V_{\sc}(x,y)
&=&V_{\sc}(x,y)-V_{\sc}(0,0)-V_{\sc}^{\rm quad}(x,y).
\end{eqnarray*}
The polynomial $V_{\sc}^{\rm quad}$ is defined from the actual beam moments and is not identified with the full physical potential of a uniform comparison beam outside that beam's boundary.
Using (\ref{kv2}), the single-particle Hamiltonian, with its constant center value removed, becomes
\begin{equation} \label{ham0p}
h_{\perp,i}
=\hal m_0c^2\beta^2\gamma
\left(x_i'^2+\kappa_x^2x_i^2+y_i'^2+\kappa_y^2y_i^2\right)
+\frac{q}{\gamma^2}\,\Delta V_{\sc}(x_i,y_i) \; .
\end{equation}
For a \kv distribution the self-field is linear throughout the occupied beam region and $\Delta V_{\sc}=0$ there.
For a non-\kv distribution the remaining potential must be included in the canonical transformation.
This case will be treated in section~\ref{non-kv}.

\subsection{The first Canonical Transformation for a \kv distribution}
\label{first_kv}
\begin{figure}[ht]
\centering\epsfig{file=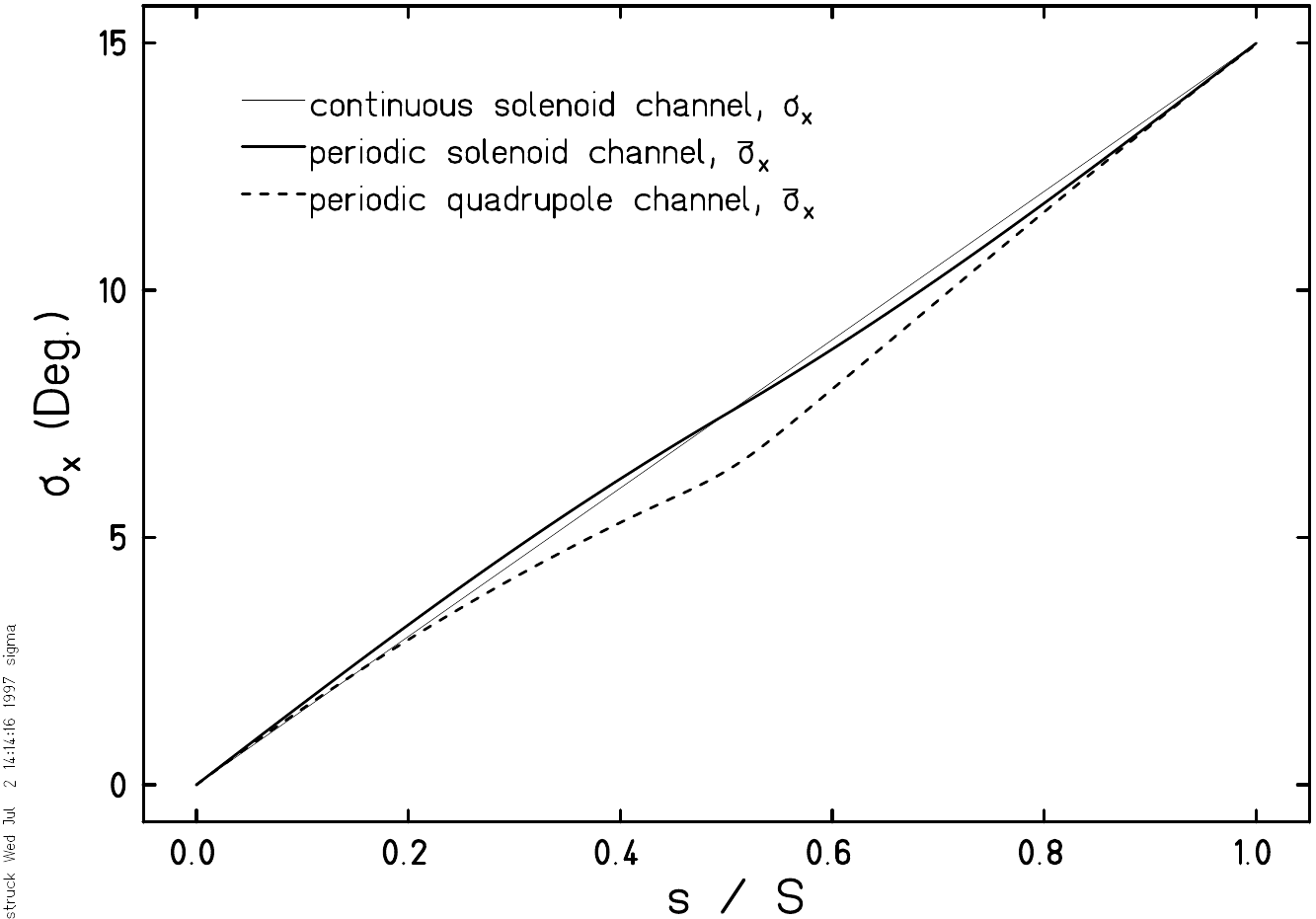,height=75mm}
\caption{Phase advance functions $\sigma_x(s)$ versus $s$ within one
focusing period at $\sigma_0=60^\circ$, $\sigma=15^\circ$.}
\label{sigma-x}
\end{figure}

For a \kv distributed beam, we define the first canonical transformation as an $s$-dependent phase displacement within the continuous focusing system.
In the $x$-plane this corresponds to a time displacement $\Delta t_x$ or a distance $\ell_x(s)=c\beta\Delta t_x$ given by
\begin{equation} \label{dx}
\ell_x(s)=\int\limits_{s_0}^{s}
\left(\frac{\<x^2\>}{\<\bar{x}^2\>(z)}-1\right)\,dz \; .
\end{equation}
The two separable linear planes may have different phase displacements.
As can be seen directly from the definitions of the phase advances of both focusing systems,
$$
\sigma_x(s)=\frac{\varepsilon_x}{\<x^2\>}\,(s-s_0),
\qquad
\bar\sigma_x(s)=\bar\varepsilon_x
\int\limits_{s_0}^{s}\frac{dz}{\<\bar{x}^2\>(z)},
$$
the phase advance of the $s$-dependent system at $s$ agrees with that of the continuous system at $s+\ell_x$,
\begin{eqnarray*}
\sigma_x(s+\ell_x)
&=&\frac{\varepsilon_x}{\<x^2\>}\,[s-s_0+\ell_x]\\
&=&\frac{\varepsilon_x}{\<x^2\>}
\int\limits_{s_0}^{s}\frac{\<x^2\>}{\<\bar{x}^2\>(z)}\,dz
=\bar\sigma_x(s),
\end{eqnarray*}
since $\varepsilon_x=\bar\varepsilon_x$ for the linear \kv dynamics.
We thus transform the particles of the continuous focusing channel to the phase corresponding to the $s$-dependent system.
The meaning of the transformation is illustrated graphically in Fig.~\ref{sigma-x}.
\begin{figure}[ht]
\centering\epsfig{file=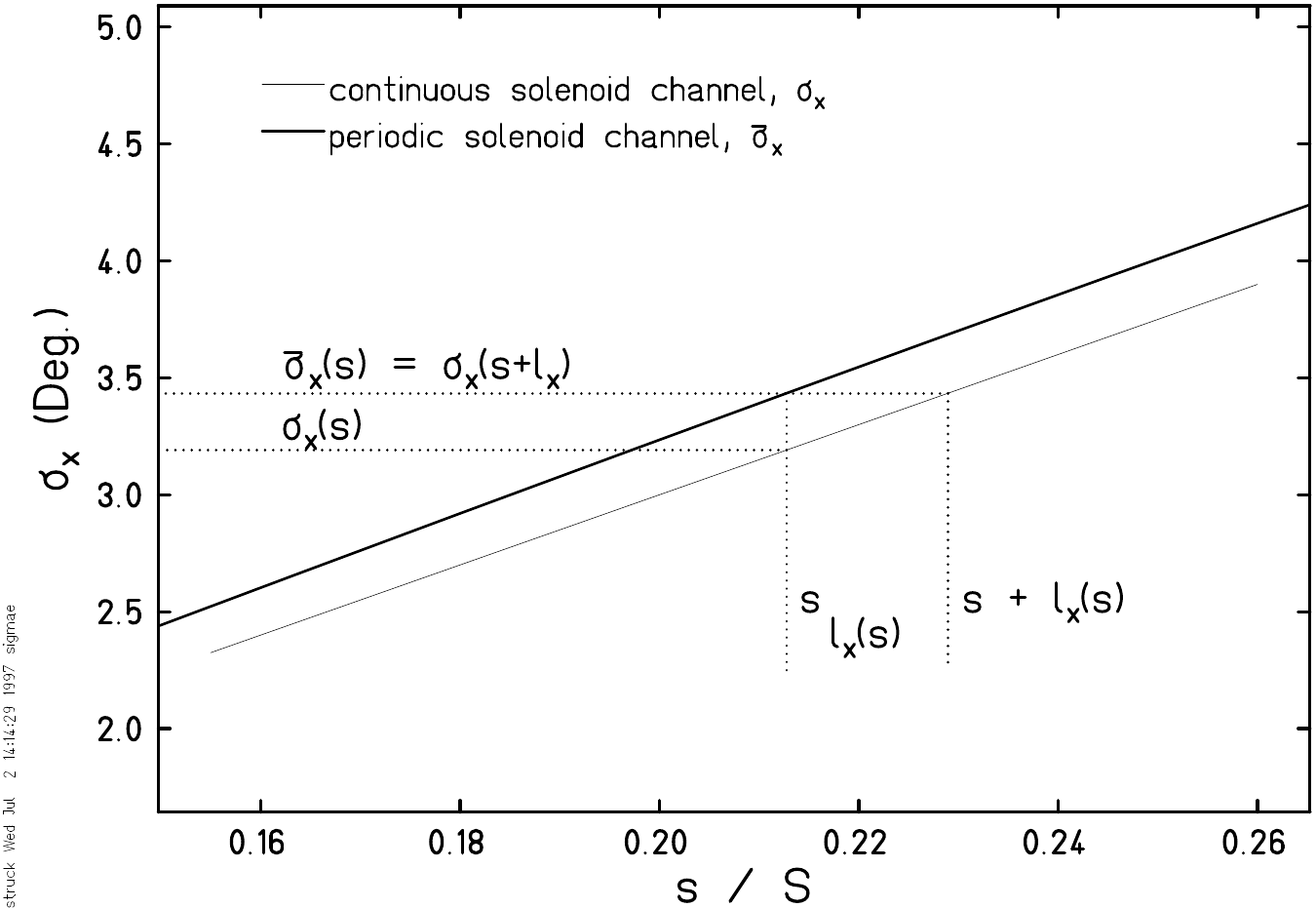,height=75mm}
\caption{Enlarged part of Fig.~\ref{sigma-x}.}
\end{figure}

The $s$-dependent generating function $F_2$ of the transformation is given by
\begin{eqnarray} \label{gen1}
F_2(x,\tilde{x}',y,\tilde{y}',s)&=&
\hal m_0c\beta\gamma\sum_i\left[
\left(\kappa_x^{-1}\tilde{x}_i'^2+\kappa_x x_i^2\right)
\tan\Psi_x(s)+\frac{2x_i\tilde{x}_i'}{\cos\Psi_x(s)}\right]
\nonumber\\
&&\mbox{}+\similar \; .
\end{eqnarray}
Here the canonical momenta are $m_0c\beta\gamma$ times the variables denoted by primes.
After the first transformation, these momentum variables need not equal derivatives of the transformed coordinates with respect to the original parameter $s$.
The abbreviation $\Psi_x(s)$ denotes the difference of the phase advances,
\begin{equation} \label{phadv}
\Psi_x(s)=\bar\sigma_x(s)-\sigma_x(s)=\kappa_x\ell_x(s) \; .
\end{equation}
The linear transformation following from (\ref{gen1}) can be written in matrix notation,
\begin{equation} \label{tra1}
\pmatrix{x_i\cr ~\cr x_i'\cr}
=\pmatrix{\cos\Psi_x(s)&-\kappa_x^{-1}\sin\Psi_x(s)\cr ~\cr
\kappa_x\sin\Psi_x(s)&\cos\Psi_x(s)\cr}
\pmatrix{\tilde{x}_i\cr ~\cr\tilde{x}_i'\cr} \; ,
\end{equation}
and likewise in $y$.
The generating function (\ref{gen1}) requires $\cos\Psi_x\ne0$, but the matrix transformation remains regular at these phase values.
The partial time derivative of (\ref{gen1}) is
$$
c\beta\frac{\partial F_2}{\partial s}
=\hal m_0c^2\beta^2\gamma
\left(\frac{\<x^2\>}{\<\bar{x}^2\>}-1\right)
\sum_i\left(\tilde{x}_i'^2+\kappa_x^2\tilde{x}_i^2\right)
+\similar \; .
$$
The transformed characteristic Hamiltonian is obtained by expressing the sum of (\ref{ham0p}) in the new coordinates and adding this derivative,
\begin{eqnarray} \label{ham1}
\tilde{\mathcal H}_{\perp}
&=&\mathcal H_{\perp}+c\beta\frac{\partial F_2}{\partial s}
\nonumber\\
&=&\hal m_0c^2\beta^2\gamma\sum_i\left[
\frac{\<x^2\>}{\<\bar{x}^2\>}
\left(\tilde{x}_i'^2+\kappa_x^2\tilde{x}_i^2\right)
+\frac{\<y^2\>}{\<\bar{y}^2\>}
\left(\tilde{y}_i'^2+\kappa_y^2\tilde{y}_i^2\right)\right] \; .
\end{eqnarray}

\subsection{The first Canonical Transformation for non-\kv distributions}
\label{non-kv}
If the remaining potential in (\ref{ham0p}) does not vanish, the single-particle equations of motion are generally nonlinear.
The linear transformation (\ref{tra1}) remains canonical, but it no longer represents the actual phase flow of the continuous beam.
An infinitesimal time displacement must be generated by the complete single-particle Hamiltonian, including its nonlinear potential.
When the two planes are coupled by that potential, a common displacement $\delta\ell$ is required.
For the phase normalization used here, this requires the similarity condition
\begin{equation} \label{simi}
\frac{\<x^2\>}{\<\bar{x}^2\>}
=\frac{\<y^2\>}{\<\bar{y}^2\>} \; ,
\end{equation}
with
$$
\delta\ell=
\left(\frac{\<x^2\>}{\<\bar{x}^2\>}-1\right)\delta s.
$$
At a fixed starting point, the type-2 generating function for this infinitesimal displacement is
\begin{eqnarray} \label{gen1p}
F_2(x,\tilde{x}',y,\tilde{y}';\delta\ell)
&=&m_0c\beta\gamma\sum_i
\left(x_i\tilde{x}_i'+y_i\tilde{y}_i'\right)
\nonumber\\
&&\mbox{}+\frac{\delta\ell}{c\beta}\,
\mathcal H_{\perp}(x,\tilde{x}',y,\tilde{y}')
+O(\delta\ell^2) \; .
\end{eqnarray}
Consequently, the transformation in the $x$-plane is
\begin{equation} \label{tra1p}
\pmatrix{x_i\cr ~\cr x_i'\cr}
=\pmatrix{1&-\delta\ell\cr ~\cr\kappa_x^2\delta\ell&1\cr}
\pmatrix{\tilde{x}_i\cr ~\cr\tilde{x}_i'\cr}
+\pmatrix{0\cr ~\cr
\displaystyle\frac{q\delta\ell}{m_0c^2\beta^2\gamma^3}
\frac{\partial\Delta V_{\sc}}{\partial\tilde{x}_i}\cr}
+O(\delta\ell^2) \; ,
\end{equation}
with the derivative evaluated at $(\tilde{x}_i,\tilde{y}_i)$, and likewise in $y$.
Keeping only the quadratic part of (\ref{gen1p}) would omit the nonlinear term in (\ref{tra1p}) already at first order.
Repeated infinitesimal linear transformations therefore cannot reproduce the missing nonlinear dynamics.

The complete autonomous Hamiltonian does generate a canonical phase flow, even when that flow has no available closed analytical expression.
For the common finite displacement $\ell(s)$, its exact characteristic Hamiltonian becomes, up to a coordinate-independent term,
\begin{equation} \label{ham1p}
\tilde{\mathcal H}_{\perp}
=\bigl[1+\ell'(s)\bigr]\,
\mathcal H_{\perp}(\tilde{x},\tilde{x}',\tilde{y},\tilde{y}')
=\frac{\<x^2\>}{\<\bar{x}^2\>}\,
\mathcal H_{\perp}(\tilde{x},\tilde{x}',\tilde{y},\tilde{y}') \; .
\end{equation}
The factor in (\ref{ham1p}) multiplies the nonlinear potential as well as the quadratic terms.
This distinction is essential when performing the second transformation.

\subsection{The second Canonical Transformation}
After correlating the phases for a \kv beam, the coordinates of both systems can be related linearly.
The same linear map can also be used to match the initial moments of a non-\kv distribution.
We write the generating function of this second canonical transformation as
\begin{eqnarray} \label{gen2}
\tilde F_2(\tilde{x},\bar{x}',\tilde{y},\bar{y}',s)
&=&m_0c\beta\gamma\sum_i\left(
\sqrt{\frac{\<\bar{x}^2\>}{\<x^2\>}}\,
\tilde{x}_i\bar{x}_i'
-\hal\frac{\<\bar{x}\bar{x}'\>}{\<x^2\>}\,
\tilde{x}_i^2\right)
\nonumber\\
&&\mbox{}+\similar \; .
\end{eqnarray}
The following linear transformation is calculated from (\ref{gen2}),
\begin{equation} \label{tra2}
\pmatrix{\tilde{x}_i\cr ~\cr\tilde{x}_i'\cr}
=\pmatrix{\sqrt{\<x^2\>/\<\bar{x}^2\>}&0\cr ~\cr
-\<\bar{x}\bar{x}'\>/\sqrt{\<x^2\>\<\bar{x}^2\>}&
\sqrt{\<\bar{x}^2\>/\<x^2\>}\cr}
\pmatrix{\bar{x}_i\cr ~\cr\bar{x}_i'\cr} \; .
\end{equation}
Its partial time derivative is
$$
c\beta\frac{\partial\tilde F_2}{\partial s}
=m_0c^2\beta^2\gamma\sum_i\left(
\frac{\<\bar{x}\bar{x}'\>}{\<\bar{x}^2\>}
\bar{x}_i\bar{x}_i'
-\frac{\bar{x}_i^2}{4\<\bar{x}^2\>}
\frac{d^2\<\bar{x}^2\>}{ds^2}\right)+\similar \; .
$$
For a \kv distribution, substitution in (\ref{ham1}) gives
\begin{eqnarray} \label{ham2}
\bar{\mathcal H}_{\perp}
&=&\tilde{\mathcal H}_{\perp}
+c\beta\frac{\partial\tilde F_2}{\partial s}
\nonumber\\
&=&\hal m_0c^2\beta^2\gamma\sum_i\left[
\bar{x}_i'^2+\bar{x}_i^2\left(
-\frac{1}{\sqrt{\<\bar{x}^2\>}}
\frac{d^2\sqrt{\<\bar{x}^2\>}}{ds^2}
+\frac{\varepsilon_x^2}{\<\bar{x}^2\>^2}\right)\right]
\nonumber\\
&&\mbox{}+\simi \; .
\end{eqnarray}
Here $\varepsilon_x$ is constant and agrees with $\bar\varepsilon_x$.
Using the envelope equation (\ref{kv1}), the focusing function of the periodic \kv channel satisfies
$$
\bar{k}_x^2(s)
=-\frac{1}{\sqrt{\<\bar{x}^2\>}}
\frac{d^2\sqrt{\<\bar{x}^2\>}}{ds^2}
+\frac{\varepsilon_x^2}{\<\bar{x}^2\>^2}
+\frac{q}{m_0c^2\beta^2\gamma^3}\,
\frac{\<\bar{x}\bar E_x\>}{\<\bar{x}^2\>} \; .
$$
Since the \kv self-field is linear inside the beam, this expression puts (\ref{ham2}) in the correct single-particle form and yields the equations of motion (\ref{bewe}).
Thus the two transformations give the exact correspondence for the \kv model, subject to a positive matched envelope.

For non-\kv distributions, the logarithmic scaling of the self-potential does not remove the extra factor in (\ref{ham1p}).
To see this explicitly, consider a round beam and set
$$
b(s)=\sqrt{\frac{\<\bar{x}^2\>(s)}{\<x^2\>}}
=\sqrt{\frac{\<\bar{y}^2\>(s)}{\<y^2\>}}.
$$
The coordinate map (\ref{tra2}) then reads
$$
\bar{x}_i=b\tilde{x}_i,
\qquad
\bar{x}_i'=\frac{\tilde{x}_i'}{b}+b'\tilde{x}_i,
$$
and likewise in $y$.
Since the reference distribution is stationary, its complete phase flow leaves its density unchanged before this scale transformation.
For a fixed total current, the continuum logarithmic kernel gives
$$
\bar V_{\sc}(\bar{x},\bar{y},s)
=V_{\sc}\left(\frac{\bar{x}}b,\frac{\bar{y}}b\right)
-\frac{2I}{c\beta}\ln b,
$$
where both potentials use the same fixed reference length.
The additive logarithm does not affect the force, but it must be retained in a comparison of field energies.
In contrast, transforming the complete characteristic Hamiltonian by (\ref{ham1p}) and (\ref{tra2}) gives the potential contribution
$$
\frac{q}{\gamma^2b^2}
V_{\sc}\left(\frac{\bar{x}}b,\frac{\bar{y}}b\right)
$$
up to coordinate-independent terms, together with a quadratic focusing term.
Its self-field force therefore differs by a factor $b^{-2}$ from that of the scaled beam at the original current.
For a quadratic self-potential this difference can be absorbed into the focusing function, as in the \kv case.
For a nonlinear self-potential it cannot generally be absorbed into a linear external focusing force.
Consequently, the canonical construction does not prove that a stationary non-\kv distribution becomes an exact periodic solution of the physical Vlasov--Poisson system.
It supplies a useful initial matching prescription, with subsequent evolution governed by the actual self-consistent field.

Both linear matrices (\ref{tra1}) and (\ref{tra2}) have unit determinant.
For any distribution, a common linear map $M_x$ in the $x,x'$ plane transforms its covariance matrix according to
$$
\bar\Sigma_x=M_x\Sigma_xM_x^{\rm T},
\qquad
\bar\varepsilon_x^2=\det\bar\Sigma_x
=(\det M_x)^2\det\Sigma_x=\varepsilon_x^2.
$$
Thus the linear matching maps preserve rms emittance at every application, not only at locations where $\ell_x=0$.
Any subsequent non-\kv emittance change results from the physical nonlinear evolution and cannot be obtained by composing these common linear maps alone.
At $\ell_x=0$, the first map reduces to the identity and the initial matching is given directly by (\ref{tra2}).
Its implication for the initial emittance derivative is considered in the next section.

Finally, the collective transverse energy must be distinguished from the characteristic generator used above.
Within the interaction model (\ref{scpot_l}) and with its fixed potential reference, the collective energy is
\begin{equation} \label{hbar}
\bar H_{\perp}
=\hal m_0c^2\beta^2\gamma\sum_{i=1}^{N}
\left(\bar{x}_i'^2+\bar{y}_i'^2
+\bar{k}_x^2(s)\bar{x}_i^2+\bar{k}_y^2(s)\bar{y}_i^2\right)
+\frac{\bar W}{\gamma^2} \; .
\end{equation}
The equations of motion and the interaction-energy balance (\ref{dw}) imply
$$
\frac{d\bar H_{\perp}}{ds}
=\hal m_0c^2\beta^2\gamma N
\left[\frac{d\bar{k}_x^2}{ds}\<\bar{x}^2\>
+\frac{d\bar{k}_y^2}{ds}\<\bar{y}^2\>\right].
$$
For a round solenoid channel this becomes
$$
\frac{d\bar H_{\perp}}{ds}
=m_0c^2\beta^2\gamma N\<\bar{x}^2\>
\frac{d\bar{k}^{\,2}}{ds},
$$
which does not vanish merely because the beam remains rotationally symmetric.
A time-dependent canonical transformation preserves the form of Hamilton's equations, but its added time derivative does not imply equality or conservation of the two physical Hamiltonians.
The emittance--field-energy relation (\ref{sr2_round}) likewise does not eliminate the work associated with the varying external focusing.

In periodic quadrupole channels, the common similarity condition (\ref{simi}) is generally not fulfilled, although the exact linear \kv construction remains applicable with separate transformations in the two planes.
For non-\kv beams the simulations presented in section~\ref{simu} permit a comparison of the practical accuracy of the matching prescription in the three channel types.
For the parameters studied there, the periodic solenoid results stay close to the continuous-channel results, whereas the quadrupole calculation shows a small further rms-emittance increase.
These observations support the usefulness of the matching procedure in the tested regime, without establishing a general conservation law or attributing every emittance change solely to the lack of rotational symmetry.
\section{INITIAL EMITTANCE GROWTH} \label{initial}
At the starting point $s_0$, we use the linear matching transformation (\ref{tra2}) to initialize a stationary continuous-channel distribution with the chosen rms envelope data.
The following argument concerns its instantaneous rms-emittance derivative.
The correlation of the particle positions is
$$
\bar{x}_i = \sqrt{\frac{\<\bar{x}^2\>}{\<x^2\>}} \cdot x_i \qquad , \qquad
\bar{y}_i = \sqrt{\frac{\<\bar{y}^2\>}{\<y^2\>}} \cdot y_i \;\;.
$$
We assume an unbunched beam with the same fixed current and particle energy in both channels, and use the free-space transverse self-field.
If at this point the similarity condition
$$
\frac{\<\bar{x}^2\>}{\<x^2\>}=\frac{\<\bar{y}^2\>}{\<y^2\>}
$$
is fulfilled, the transformed electric space charge fields at
$s_0$ can be correlated to the original ones according to (\ref{e_x_l}):
$$
E_x(x_i,y_i) = \sqrt{\frac{\<\bar{x}^2\>}{\<x^2\>}} \cdot
\bar{E}_x\bigl(\bar{x}_i,\bar{y}_i\bigr) \; .
$$
The stationary reference beam has $\<xx'\>=0$.
We can now express $x_i'E_x(x_i,y_i)$ at $s_0$ in terms of the barred coordinates and average over all particles
\begin{equation} \label{epsdericorr}
\<x^2\>\<x'E_x\> \, = \, \<\bar{x}^2\>\<\bar{x}'\bar{E}_x\> -
\<\bar{x}\bar{x}'\>\<\bar{x}\bar{E}_x\> \; .
\end{equation}
Due to the similarity, the same correspondence holds for the $y$-direction.
Using Eqs.~(\ref{epsderi}) and (\ref{epsdericorr}), we can correlate the change of the rms-emittances in the continuous channel to the corresponding change in the periodic channel
$$
\frac{d}{ds}\, \varepsilon_x^2\, (s_0) = \frac{2q}{m_0 c^2 \beta^2 \gamma^3}
\<x^2\>\<x'E_x\> = \frac{d}{ds} \, \bar{\varepsilon}_x^2\, (s_0) \; .
$$
In other words, the matching transformation (\ref{tra2}) does not
change the gradient of the rms-emittance function at $s_0$.
If we transform a stationary continuous-channel distribution, its reference rms-emittance is constant and we obtain
$$
\frac{d}{ds}\, \bar{\varepsilon}_x^2\bigl(s_0\bigr) = 0 \;\; .
$$
The matching transformation therefore preserves a vanishing initial rms-emittance derivative.
A nonzero quadratic or higher-order change after $s_0$ remains possible.
For example, consider the common scaling $\bar{x}=bx$, $\bar{y}=by$, $\bar{x}'=x'/b$, and $\bar{y}'=y'/b$ at a point with zero envelope slopes, where $b>0$.
If the initial external focusing is chosen as
$$
\bar{k}_x^2(s_0)=\frac{\kappa_x^2}{b^4}
+\frac{k_0^2-\kappa_x^2}{b^2}\; ,
$$
and likewise in $y$, the initial second derivative of the rms radius also vanishes by (\ref{kv1}).
For a round stationary reference distribution $f=f(w)$, the initial emittance curvature nevertheless satisfies
\begin{eqnarray} \label{initialcurvature}
\left.\frac{d^2\bar{\varepsilon}_x^2}{ds^2}\right|_{s_0}
&=&2\left(\frac{q}{m_0c^2\beta^2\gamma^3}\right)^2
\<x^2\>\left(1-\frac{1}{b^2}\right) \nonumber \\
&&\mbox{}\times\left\langle
\left(E_x-\frac{\<xE_x\>}{\<x^2\>}x\right)^2
\right\rangle \; .
\end{eqnarray}
All unbarred moments in this expression refer to the stationary distribution at launch.
This result follows by differentiating (\ref{epsderi}) and using the stationary moment identities together with the initially vanishing transverse charge current.
The curvature is generally nonzero for a nonlinear space charge field and $b\ne1$.
It vanishes for the linear field of a \kv beam, consistently with its exact linear matching theory.
\section{MAXIMUM EMITTANCE GROWTH FOR NON-STATIONARY\\ DISTRIBUTIONS} \label{final}
For a round unbunched beam, Eq.~(\ref{sr2_round}) can be rewritten in the following form
\begin{equation} \label{sr3}
\frac{1}{\<x^2\>(s)}\, \frac{d\, \varepsilon_x^2(s)}{ds} +
\frac{1}{\<y^2\>(s)}\, \frac{d\, \varepsilon_y^2(s)}{ds} +
\frac{K}{w_0}\cdot \frac{d}{ds}\bigl[ W(s)-W_u(s)\bigr] = 0 \; ,
\end{equation}
with $K$ denoting the ``generalized perveance'' (\ref{perve})
and $w_0$ the field energy normalization factor.
For unbunched beams, the latter can be expressed as
$$
w_0 = \frac{q I N}{c\beta} = \frac{q^2 N^2}{L} \; ,
$$
Here $N$ denotes the number of particles in a fixed beam segment of length $L$, and $W$ and $W_u$ are the field energies of that segment.
If $\<x^2\>$ and $\<y^2\>$ vary little during the redistribution, we can approximate them by their initial values when integrating Eq.~(\ref{sr3}).
This gives the approximate correlation
$$
\frac{\Delta \varepsilon_x^2(s)}{\<x^2\>} +
\frac{\Delta \varepsilon_y^2(s)}{\<y^2\>} +
K\cdot \frac{\Delta\bigl( W(s) - W_u(s) \bigr) }{w_0} \simeq 0 \; .
$$
The second moments in this integrated expression are evaluated at $s=0$.
For a round beam with equal transverse emittances and $\Delta\varepsilon_x^2(s)=\varepsilon_x^2(s)-\varepsilon_x^2(0)$, it gives
$$
\frac{\varepsilon_x(s)}{\varepsilon_x(0)} \simeq \sqrt{ 1 -
\frac{K \<x^2\>}{2\varepsilon_x^2(0)}\cdot
\frac{\Delta\bigl( W(s)-W_u(s)\bigr) }{w_0} } \; .
$$
The constant-radius treatment is an approximation during emittance growth.
For fixed continuous focusing and current, an exactly constant radius in (\ref{kv1}) would also require a constant emittance.
If the radius changes appreciably, Eq.~(\ref{sr3}) must instead be integrated with the actual varying second moments.
The factor $K\<x^2\>/ \varepsilon_x^2(0)$ can be converted into a more
transparent form, if we use the transverse particle oscillation
frequencies with and without consideration of the space charge forces (``tunes'').
These oscillation frequencies are valid for linear equations of motion,
i.e.\ for a \kv beam.
As can be concluded from Eq.~(\ref{tune}),
the space charge forces cause the zero current
tune $k_0$ to be depressed down to $k$ according to
$$
\frac{k_0^2}{k^2} - 1 = \frac{K}{4 \<x^2\> k^2}\; .
$$
To define an equivalent tune for a non-\kv round beam, the reference \kv radius $a$ in (\ref{tune}) is replaced by $a^2=4\<x^2\>$ at launch.
In this context, $k$ denotes the depressed betatron frequency
the beam particles would have, if the entire particle phase space
distribution was of the \kv type.
We can use this notation for non-\kv beams also, if we keep in mind
that $k$ no longer means the actual single particle oscillation frequency,
but only a global measure for the strength of the space charge effects.
For the initially rms-matched continuous reference beam, the depressed tune $k\equiv\kappa_x$ is correlated to the initial beam moments as in Eq.~(\ref{kv2})
$$
k^2 = \frac{\<x'^2\>}{\<x^2\>} = \frac{\varepsilon_x^2(0)}{\<x^2\>^2}\; ,
$$
thus
\begin{equation} \label{egdw}
\frac{\varepsilon_x(s)}{\varepsilon_x(0)} \simeq \sqrt{
1 - 2\left( \frac{k_0^2}{k^2} - 1 \right) \cdot
\frac{\Delta\bigl( W(s)-W_u(s)\bigr) }{w_0} } \; .
\end{equation}
Among round charge distributions of the same total charge and rms radius, the uniform density has the minimum field energy when a common potential reference is used\cite{host}.
The nonnegative difference $W-W_u$ is therefore the field energy available from the initial density nonuniformity at fixed second moments.
Envelope mismatch and work by time-dependent focusing provide additional energy sources and require separate treatment.
An upper estimate for growth from the initial density nonuniformity is obtained, within the approximation above, by assuming that the entire initial excess field energy is released.
Neglecting the remaining excess field energy of the final state gives
$$
\Delta\bigl( W(s) - W_u(s)\bigr) = W(s) - W_u(s) - W(0) + W_u(0)
\approx -\bigl( W(0) - W_u(0) \bigr)\; .
$$
We thus obtain the following simple full-conversion estimate\cite{strekla,reiser}, denoted by $\varepsilon_x^{\mbx}$ within the constant-radius approximation
\begin{equation} \label{eg-formula}
\frac{\varepsilon_x^{\mbx}}{\varepsilon_x(0)} = \sqrt{ 1 +
2 f \left( \frac{k_0^2}{k^2} - 1 \right) }\; ,
\end{equation}
with the dimensionless geometry factor $f$ as a measure for
the initial inhomogeneity of the charge distribution
$$
f = \frac{W(0) - W_u(0)}{w_0}\; .
$$
Eq.~(\ref{eg-formula}) leads to good estimates especially in the high
current regime, where equilibrium charge densities are nearly uniform,
as shown in section~\ref{water}.
For the non-stationary ``water bag'' distribution, we have
$f = \frac{5}{24} - \frac{1}{2}\ln\frac{3}{2}$ as calculated in
Ref.~\cite{strekla}.
In our simulation examples, we adjusted the zero current phase advance to
$\sigma_0=k_0S=60^\circ$ and the equivalent depressed phase advance to $\sigma=kS=15^\circ$ in the continuous reference channel.
Here $S$ is the comparison period, and the corresponding periodic channels are specified by their phase advances over one cell.
Inserting these values into (\ref{eg-formula}) gives the estimate for complete release of the initial excess field energy
$$
\frac{\varepsilon_x^{\mbx}}{\varepsilon_x(0)} =
\sqrt{1 + \left(\frac{5}{12} - \ln{\frac{3}{2}} \right)
\left( \frac{60^2}{15^2} - 1\right) } \approx 1.081 \; .
$$
The emittance increase reported in Fig.~\ref{zcwbegf} is close to this estimate.
The changing rms dimensions in the simulations give corrections to the constant-radius approximation.
Accordingly, (\ref{eg-formula}) estimates growth from the initial density nonuniformity and is not a rigorous upper bound for an arbitrary periodically focused beam.
\section{RESULTS OF NUMERICAL SIMULATIONS} \label{simu}
\subsection{Overview}
\begin{figure}[hp]
\vspace*{20mm}
\begin{flushleft}
(a)
\end{flushleft}
\vspace*{-35mm}
\centering\epsfig{file=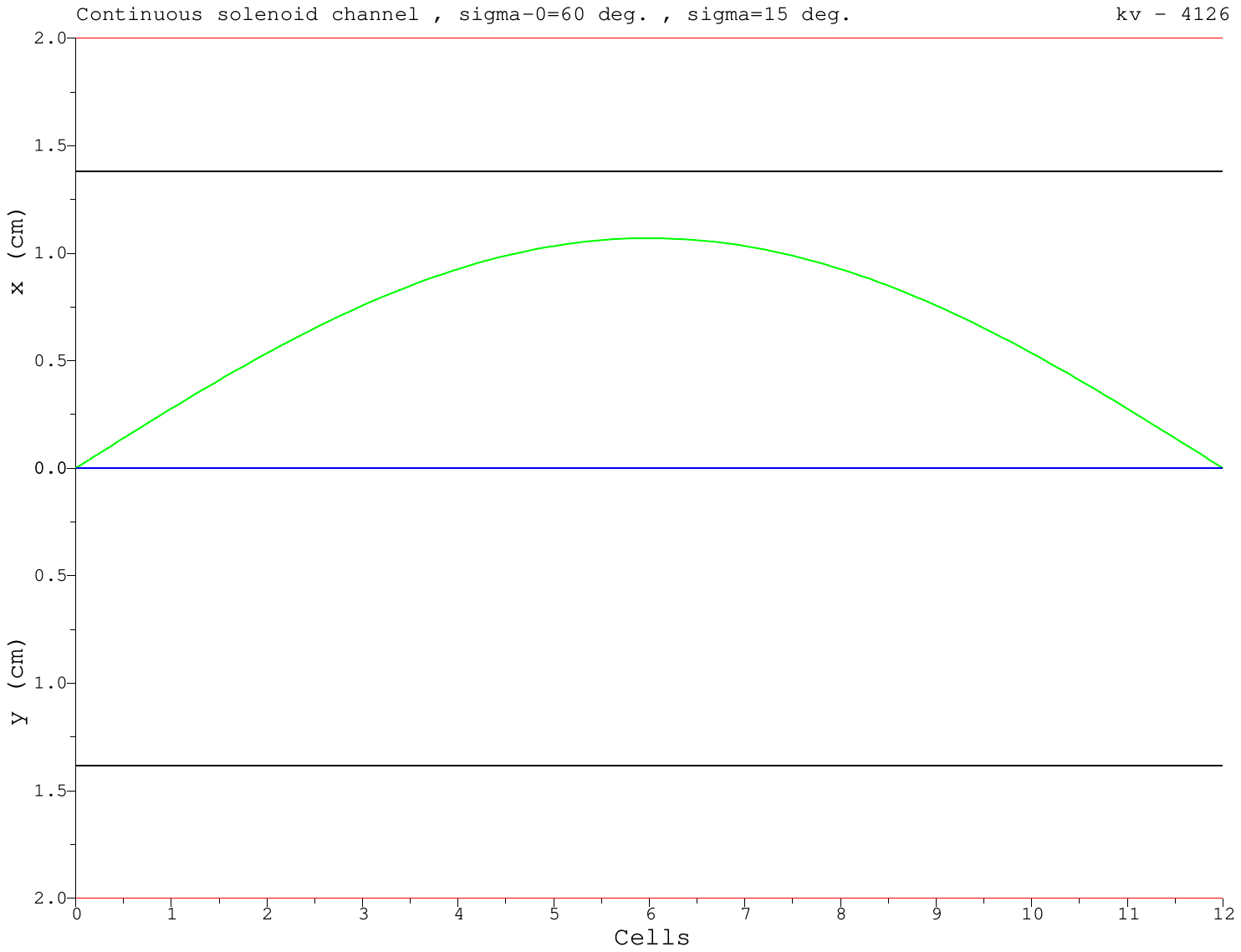,height=80mm,angle=-90}
\vspace*{20mm}
\begin{flushleft}
(b)
\end{flushleft}
\vspace*{-35mm}
\centering\epsfig{file=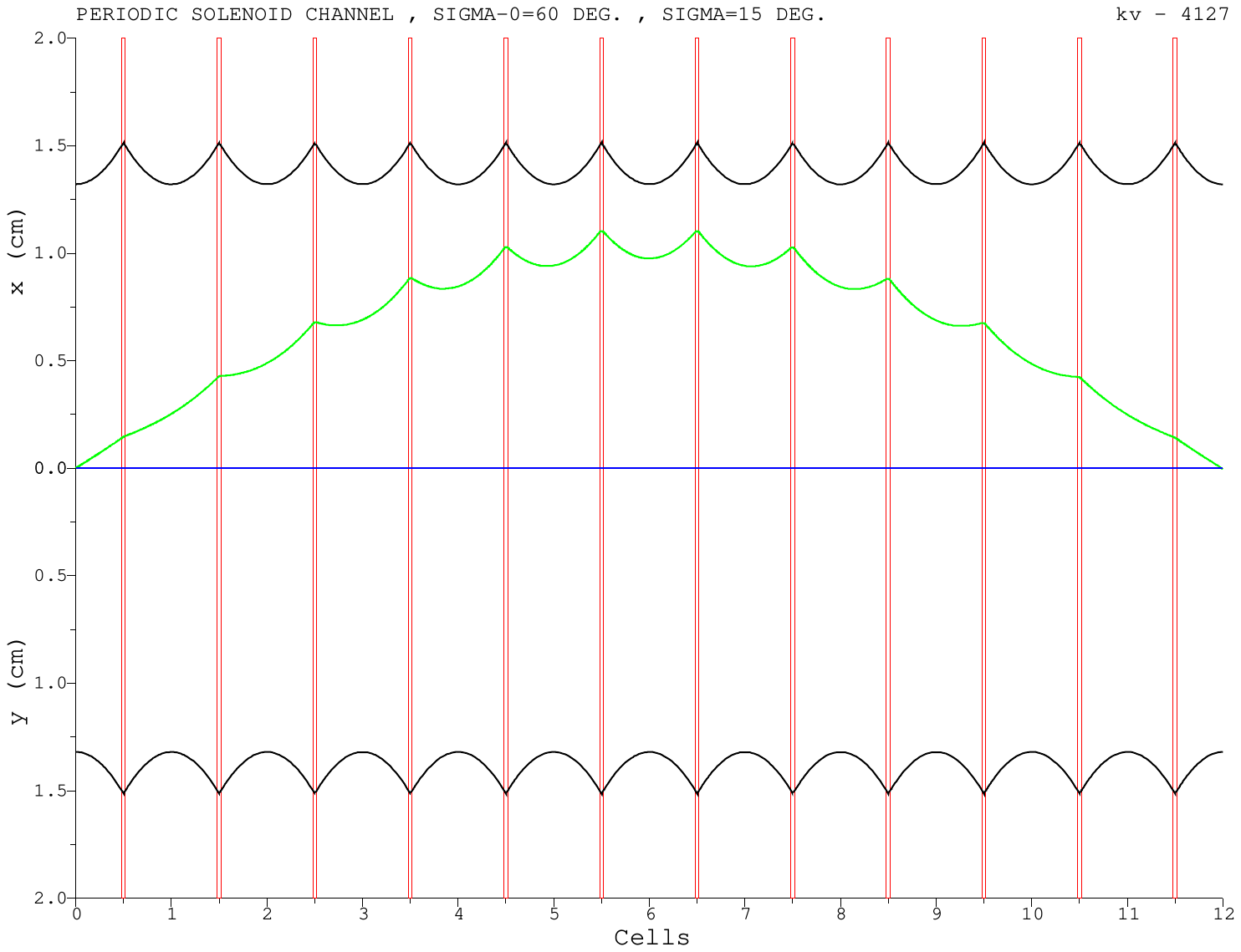,height=80mm,angle=-90}
\vspace*{20mm}
\begin{flushleft}
(c)
\end{flushleft}
\vspace*{-35mm}
\centering\epsfig{file=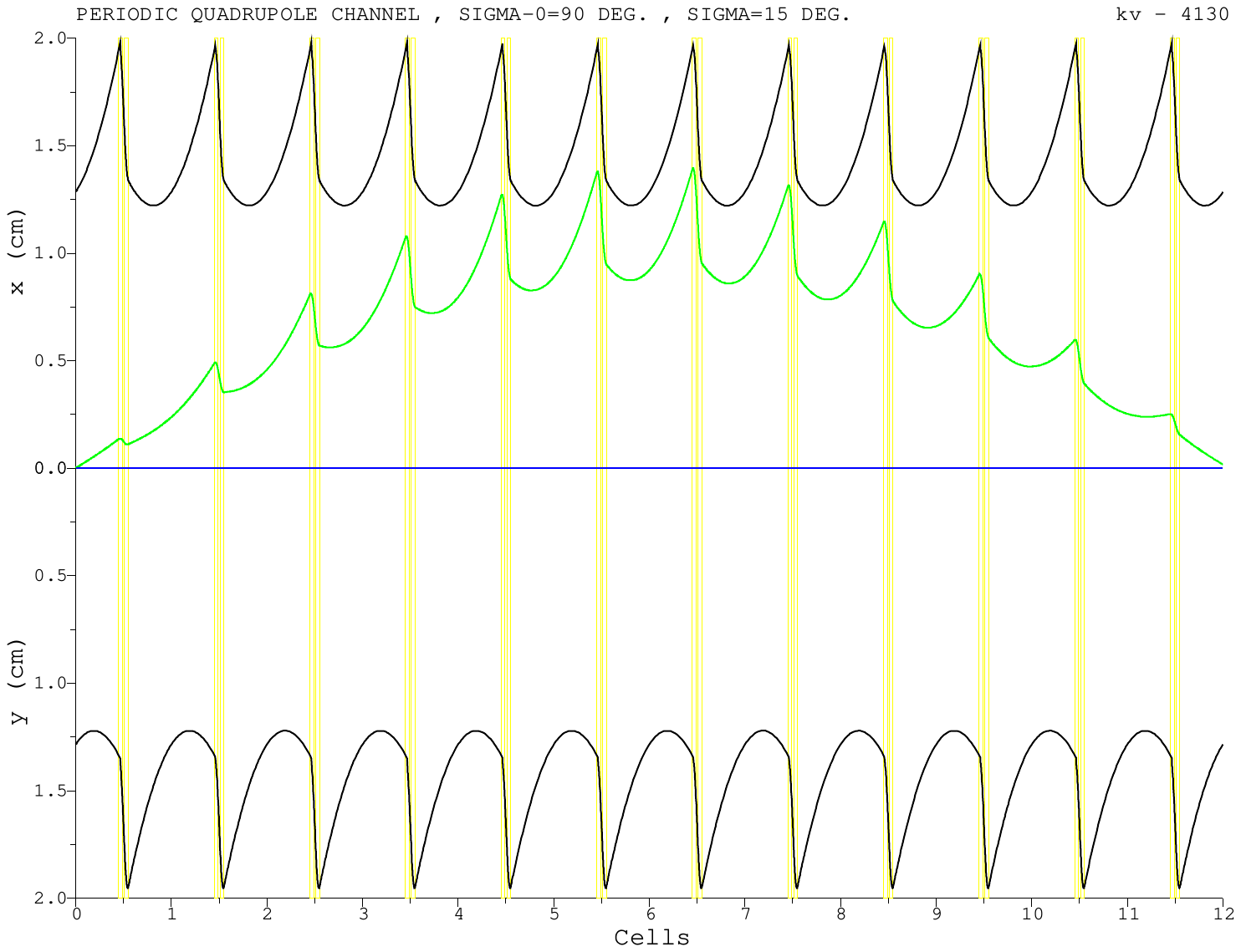,height=80mm,angle=-90}
\vspace*{-4mm}
\caption{Envelopes (black) and single particle trajectories (green) of matched \kv beams
passing through a Continuous Solenoid Channel (a), a Periodic Solenoid
Channel (b), and a Periodic Quadrupole Channel (c),
at $\sigma_0 = 60^\circ$, $\sigma = 15^\circ$.}
\label{systems}
\end{figure}

We started our simulations of charged particle beams
with three different types of initial phase space distributions:
\begin{itemize}
\item \kv distribution as described in section~\ref{kv},
\item stationary ``water bag'' distribution
as described in section~\ref{water},
\item ``water bag'' distribution generated for the zero current limit.
In consequence, it is \emph{not} stationary for finite beam currents.
\end{itemize}
As transport devices we used the three basic types, i.e.\
\begin{itemize}
\item a continuous focusing solenoid channel (Fig.~\ref{systems}a),
\item a periodic focusing solenoid channel (Fig.~\ref{systems}b),
\item a periodic focusing quadrupole channel (Fig.~\ref{systems}c).
\end{itemize}
In all three channels, the focusing strength has been adjusted to generate
a zero current phase advance of $\sigma_0 = 60^\circ$ over the
distance of one focusing period $S$.
The beam current and the initial value of the rms-emittance have been
chosen equally for all three types of distributions
to yield an equivalent K-V depressed phase advance of $\sigma = 15^\circ$.
For the distributions with nonlinear self-fields, this is a measure of
space-charge strength based on the matched second moments and does not
denote a common single-particle oscillation frequency.
The numerical results below describe the parameter sets and propagation
distances investigated.
In particular, interpreting a small persistent emittance slope as a physical
effect requires convergence checks with respect to particle number,
field resolution, integration step size, and the initial sampling.
Such checks are necessary to separate collective dynamics from numerical
diffusion and focusing-dependent grid effects\cite{hofnoise}.

\begin{figure}
\centering\epsfig{file=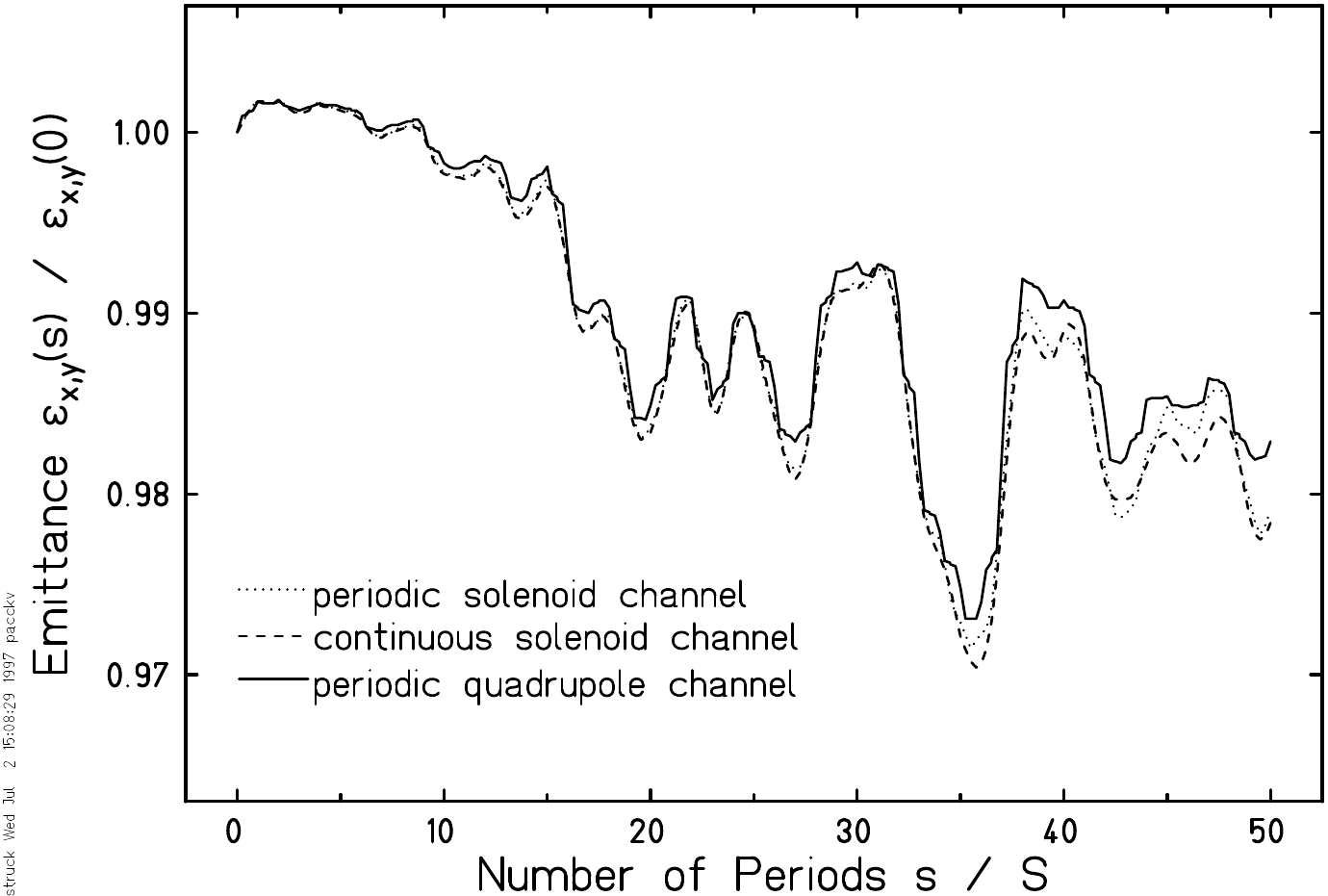,height=75mm}
\caption{Emittance growth factors versus the number of periods for an
initial \kv distribution at $\sigma_0 = 60^\circ$, $\sigma = 15^\circ$.}
\label{kvegf}
\end{figure}

\begin{figure}
\centering\epsfig{file=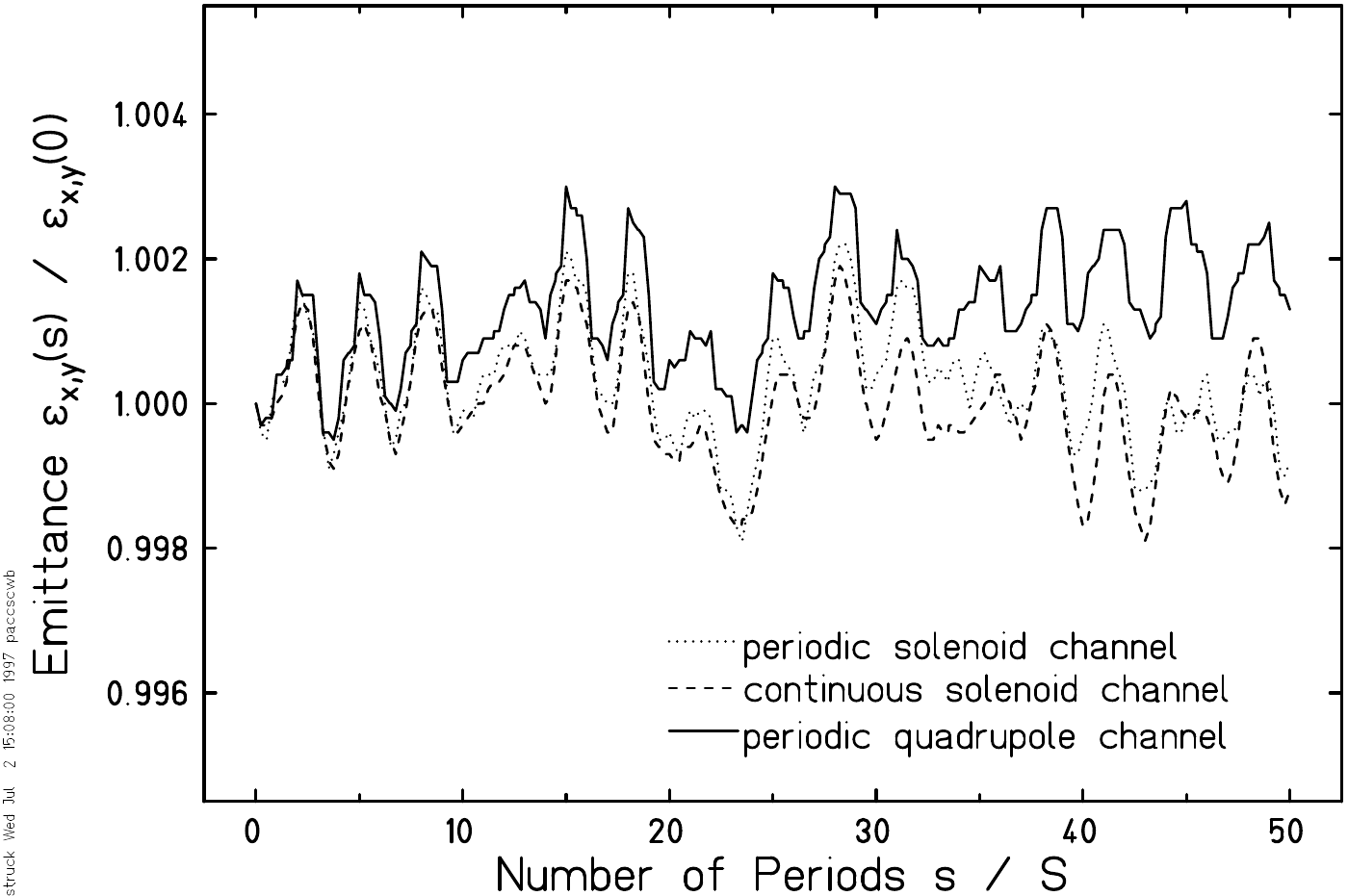,height=75mm}
\caption{Emittance growth factors versus the number of periods for a
stationary ``water bag'' distribution at $\sigma_0 = 60^\circ$,
$\sigma = 15^\circ$.}
\label{scwbegf}
\end{figure}

Figures~\ref{kvegf}, \ref{scwbegf}, and \ref{zcwbegf} show
the emittance growth factors as they occur while the respective
beam passes through the various types of transport channels.
To illustrate the behavior of the different particle phase space
distributions, two-dimensional projections of the phase space points
are plotted at specific positions $s$ along the beam line.
Since these plots look nearly identical for all three types of
channels, only the plots pertaining to the
periodic quadrupole channel are being displayed.
\subsection{``K-V'' Distribution}
In Fig.~\ref{kvegf}, the ratios
$\varepsilon_{x,y}(s)/\varepsilon_{x,y}(0)$
are plotted for an initial \kv distribution.
Since the ideal \kv distribution is an exact matched solution in all three
channels, its unperturbed evolution has constant rms-emittance.
The slight increase of the emittance just after launching the beam is
consistent with perturbations introduced by the finite number of simulation
particles ($10^5$) and the numerical field calculation.
The subsequent decrease of the rms-emittance beyond cell no.~$7$ is
associated with the redistribution described below.
As can be observed in the top picture of Fig.~\ref{wbperi}, the \kv distribution is not
sustained.
The $x,x'$- and the $y,y'$-phase space projections become rectangular
rather than elliptical.
In the $x,y$-plane, the charges rearrange to form a uniform density
in the beam center and a decreasing density at the beam edge.
This behavior is consistent with excitation of an instability of the
K-V equilibrium in the parameter regime considered.
The resulting distribution has screening properties and appears more
stationary over the propagation interval displayed.
The observations alone do not establish its stability against arbitrary perturbations.
The resulting charge distribution is less uniform than the
strictly uniform \kv state.
This transition is associated with an increase of the --- initially
vanishing --- excess field energy.
For nearly constant rms radii, equation (\ref{egdw}) relates such an increase
to a \emph{decrease} of the rms-emittance.
In a periodic channel the same interpretation is an approximation,
subject to the restrictions stated in section~\ref{final}.
We observe that the emittance growth factors agree
remarkably for all types of transport channels.
This indicates that the three channels produce similar rms evolution
in this example, consistent with the practical matching correspondence
discussed in section~\ref{catra}.
\subsection{Stationary ``Water Bag'' Distribution}
Fig.~\ref{scwbegf} shows the emittance ratios for the
stationary ``water bag'' case.
In the first step, the phase space distribution is generated
for the continuous focusing channel.
It is characterized by a uniform density of phase space points
within the boundary given by Eq.~(\ref{border}).
The space charge parameter contained in (\ref{border})
evaluates to $\kappa a = 8.19$ for our simulation examples.
This distribution can be transported directly through the
continuous solenoid channel.
For both types of periodic channels, we have to apply the matching
transformation (\ref{tra2}) before starting
the beam transport simulations.
The second order beam moments are thus adjusted to the initial moments
evaluated earlier for the \kv distribution that proved to yield
$S$-periodic (``matched'') solutions of the envelope equation (\ref{kv1}).

Regarding Fig.~\ref{scwbegf}, we observe practically no changes
of the rms-emittance values for all three types of transport channels.
For the exact stationary distribution in the continuous solenoid channel,
the emittance is a constant of motion.
Residual fluctuations in its numerical realization can arise from
finite-particle sampling and numerical errors, which also perturb the equilibrium.
The range of these fluctuations agrees for all types of transport channels.
A slight increase of the rms-emittance is observed for the quadrupole channel.
Nevertheless the increase is fairly low -- only a factor of $1.002$
is obtained after $50$ periods in our example.
The varying ellipticity and nonlinear self-field in a quadrupole channel
can contribute to departures from the initially matched distribution.
Numerical effects can also contribute to an increase of this magnitude.
The observations therefore support an approximate matching correspondence,
as discussed in section~\ref{catra}, without establishing a unique cause
of the residual emittance slope.

\begin{figure}[hp]
\begin{center}
\epsfig{file=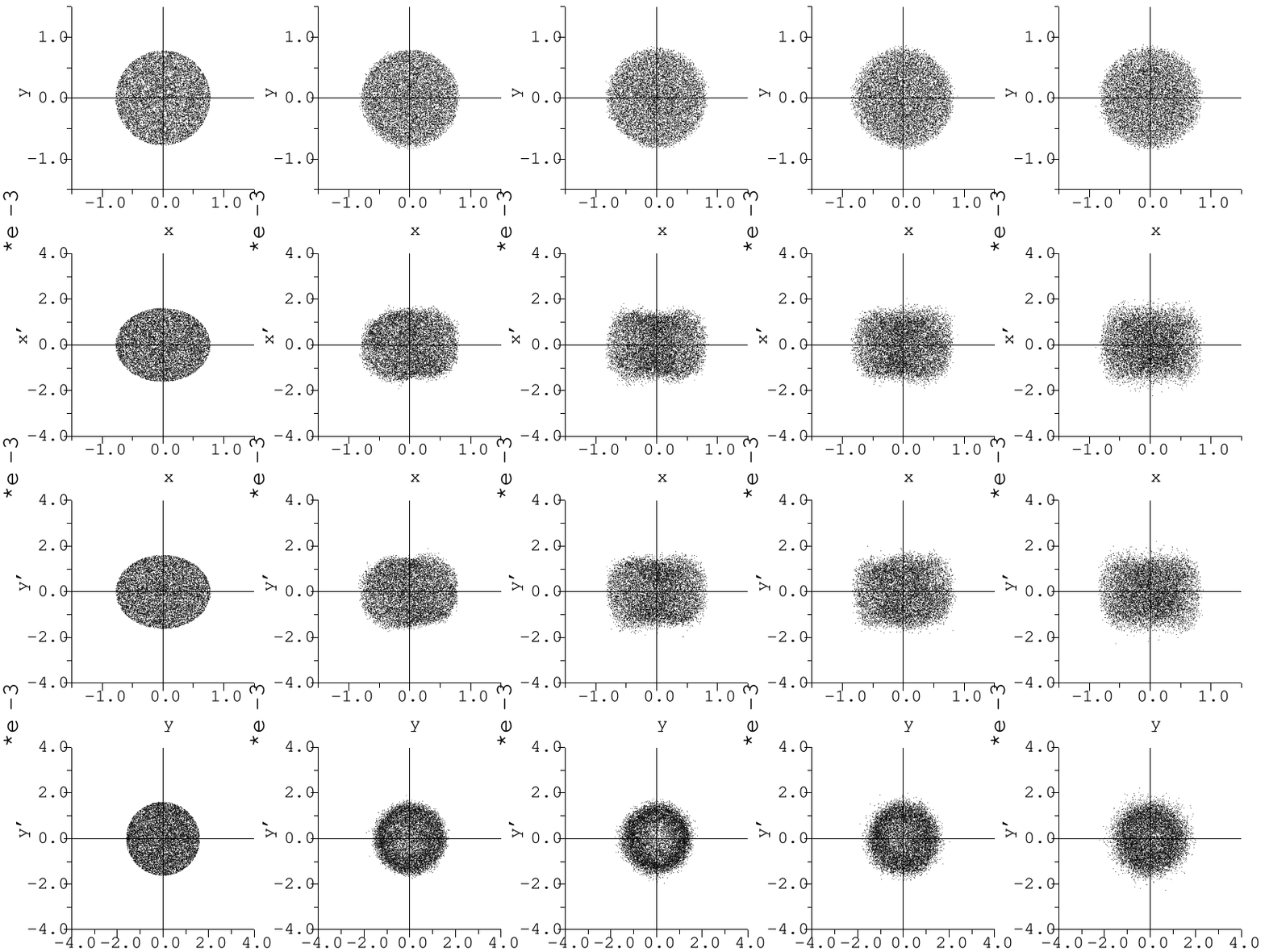,width=0.75\linewidth,angle=-90}
\\~\\ Period\hspace*{5mm} 0\hspace*{2cm} 10\hspace*{2cm} 20\hspace*{2cm} 30\hspace*{2cm} 50\hspace*{\fill}
\end{center}
\begin{center}
\epsfig{file=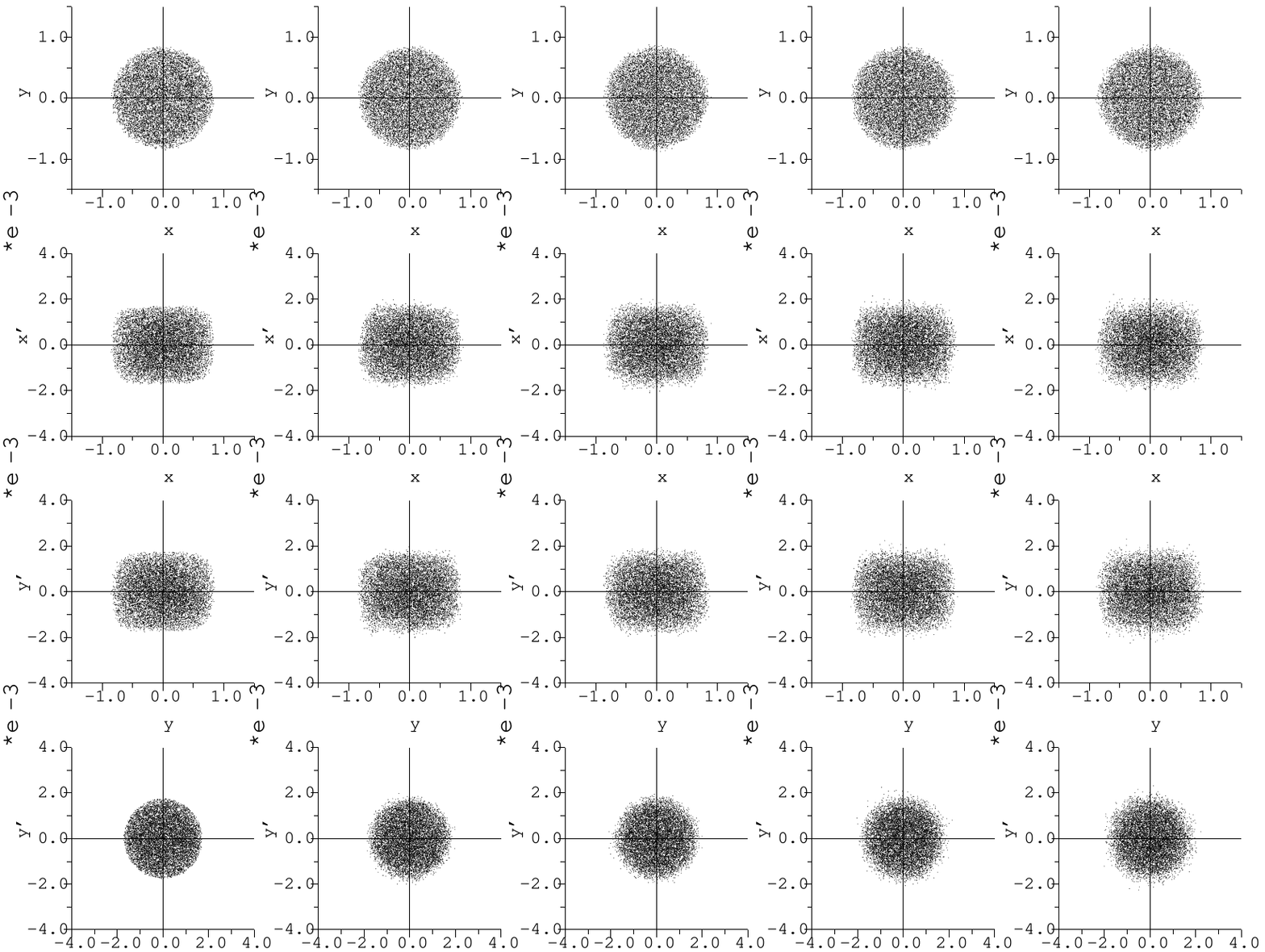,width=0.75\linewidth,angle=-90}
\\~\\ Period\hspace*{5mm} 0\hspace*{2cm} 10\hspace*{2cm} 20\hspace*{2cm} 30\hspace*{2cm} 50\hspace*{\fill}
\caption{Top: \kv distribution transformed through
a periodic quad\-ru\-pole channel at $\sigma_0 = 60^\circ$, $\sigma = 15^\circ$.
The $x,x'$- and $y,y'$-phase space projections have been transformed
to principal axes (units~: mm and mrad).\\
Bottom: Stationary ``water bag'' distribution transformed through
a periodic quadrupole channel at $\sigma_0 = 60^\circ$, $\sigma = 15^\circ$.
The $x,x'$- and $y,y'$-phase space projections have been transformed
to principal axes (units~: mm and mrad).}
\label{wbperi}
\end{center}
\end{figure}

If we examine how the phase space projections evolve in the bottom picture of Fig.~\ref{wbperi},
practically no changes in the real space plane can be observed.
On the other hand, we see that the sharp boundary in the $x',y'$-plane
is not sustained.
The ``water bag'' distribution (\ref{wbdef}) has a sharp boundary at
$w=w_{\mbx}$, which is preserved by its exact stationary Vlasov evolution.
In the numerical realization, sampling and field errors can smear the
projected boundary, while departures from exact matching in a periodic
channel can also deform it.
These effects must be distinguished from an intrinsic relaxation of an
exact continuous-channel equilibrium.
\begin{figure}
\centering\epsfig{file=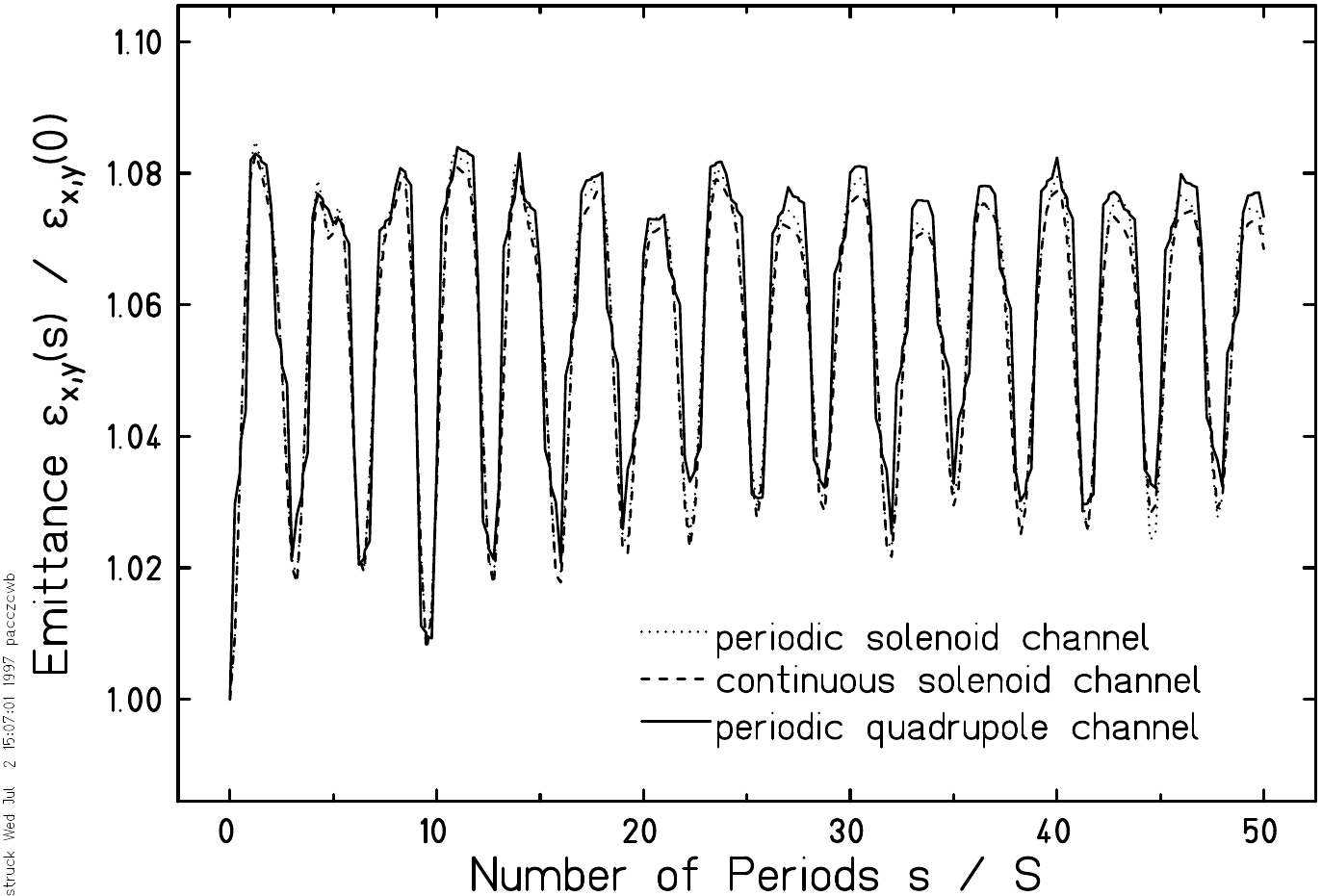,height=75mm}
\caption{Emittance growth factors versus the number of periods for a
non-stationary ``water bag'' distribution at $\sigma_0 = 60^\circ$,
$\sigma = 15^\circ$.}
\label{zcwbegf}
\end{figure}
\subsection{Non-stationary ``Water Bag'' Distribution}
\begin{figure}
\begin{center}
\epsfig{file=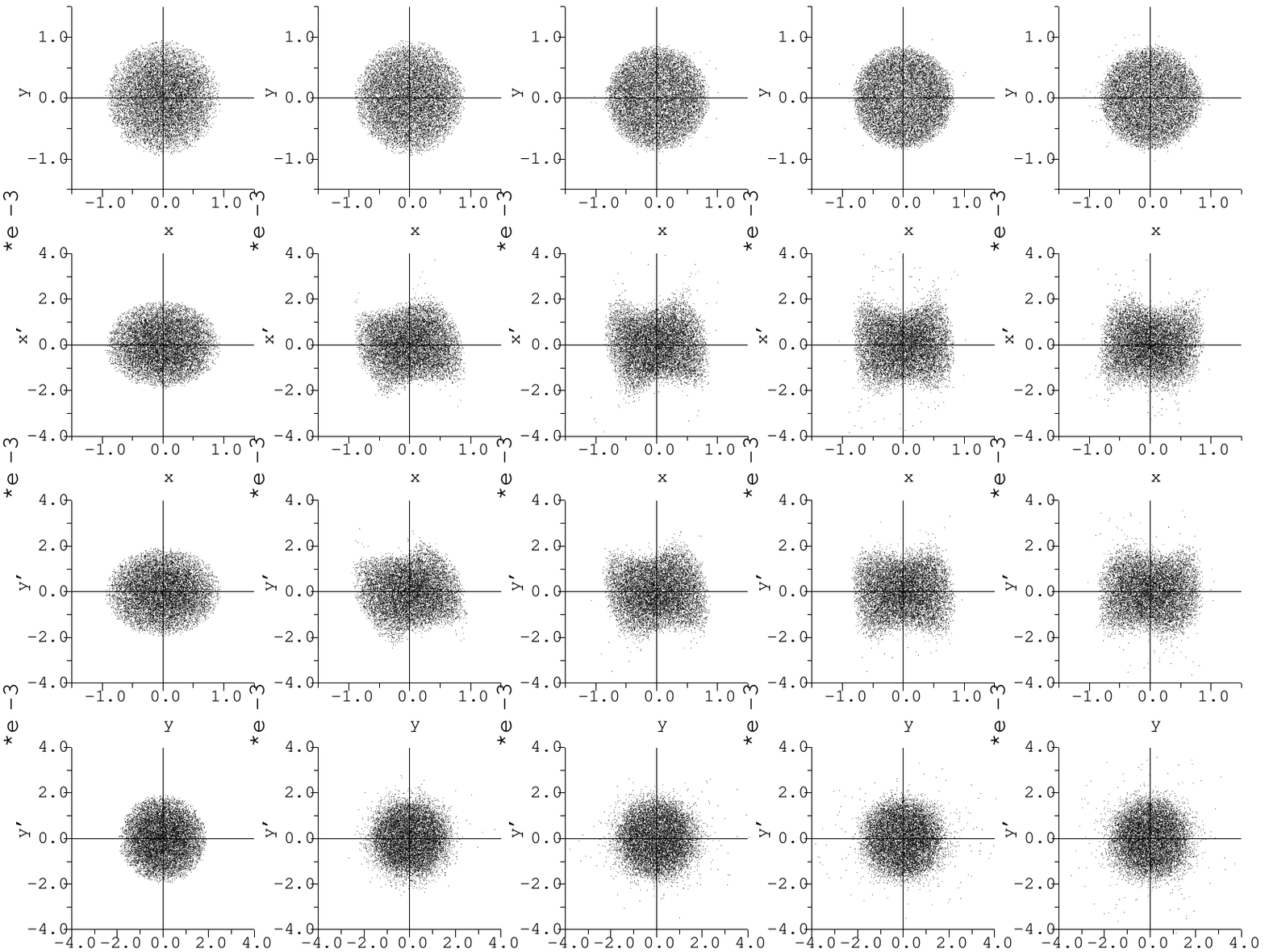,width=0.75\linewidth,angle=-90}
\\~\\ Period\hspace*{5mm} 0\hspace*{2cm} 10\hspace*{2cm} 20\hspace*{2cm} 30\hspace*{2cm} 50\hspace*{\fill}
\caption{Non-stationary ``water bag'' distribution transformed through
a periodic quadrupole channel at $\sigma_0 = 60^\circ$, $\sigma = 15^\circ$.
The $x,x'$- and $y,y'$-phase space projections have been transformed
to principal axes (units~: mm and mrad).}
\label{zcwbperi}
\end{center}
\end{figure}
In the zero current limit, the stationary ``water bag'' distribution
is characterized by an elliptical phase space boundary, and a charge
density in real space decreasing quadratically:
$n(r) \propto 1-r^2/r^2_{\mbx}$,
as has been shown in section~\ref{water}.
If this distribution is launched under space charge conditions,
it is in a non-equilibrium state, even if the second order beam
moments are perfectly matched.
This results in a complete rearrangement of the whole
populated phase space within the first plasma period.
In real space, this leads to a more uniform charge density profile,
which means that the excess field energy $W-W_u$ is reduced.
For nearly constant rms radii, Eq.~(\ref{egdw}) relates this reduction
to an \emph{increase} of the transverse rms-emittances.
If we assume the initial excess field energy to be converted completely
into transverse thermal energy, we obtain the maximum-growth estimate
within this approximation.
As already calculated in section~\ref{final}, this increase
amounts to $\approx 8 \,\%$ for the non-stationary ``water bag''
distribution transformed under space charge conditions that
cause a tune depression of $k_0/k = 4$.
The simulation examples in Fig.~\ref{zcwbegf} are consistent with this estimate.
Again, the emittance growth factors
agree for all three types of transport channels.
The channels therefore show approximately equivalent rms evolution
for this non-stationary initial distribution and the parameters investigated.

The reordering of all phase space points that occurs if the initial state
of the distribution is not stationary can be examined in Fig.~\ref{zcwbperi}.
The real space $x,y$-projection takes on a uniform density, except at the beam edge.
In the $x,x'$- and the $y,y'$-subspaces, we see that
the initial elliptical symmetry becomes rectangular,
although oscillations have not completely relaxed
at the end of our calculations (cell number~$50$).
The $x',y'$-plots show that a considerable heating takes place, which
reflects the effect that excess field energy is converted into
transverse thermal energy.
\subsection{Long Distance Beam Transport}
\begin{figure}
\centering\epsfig{file=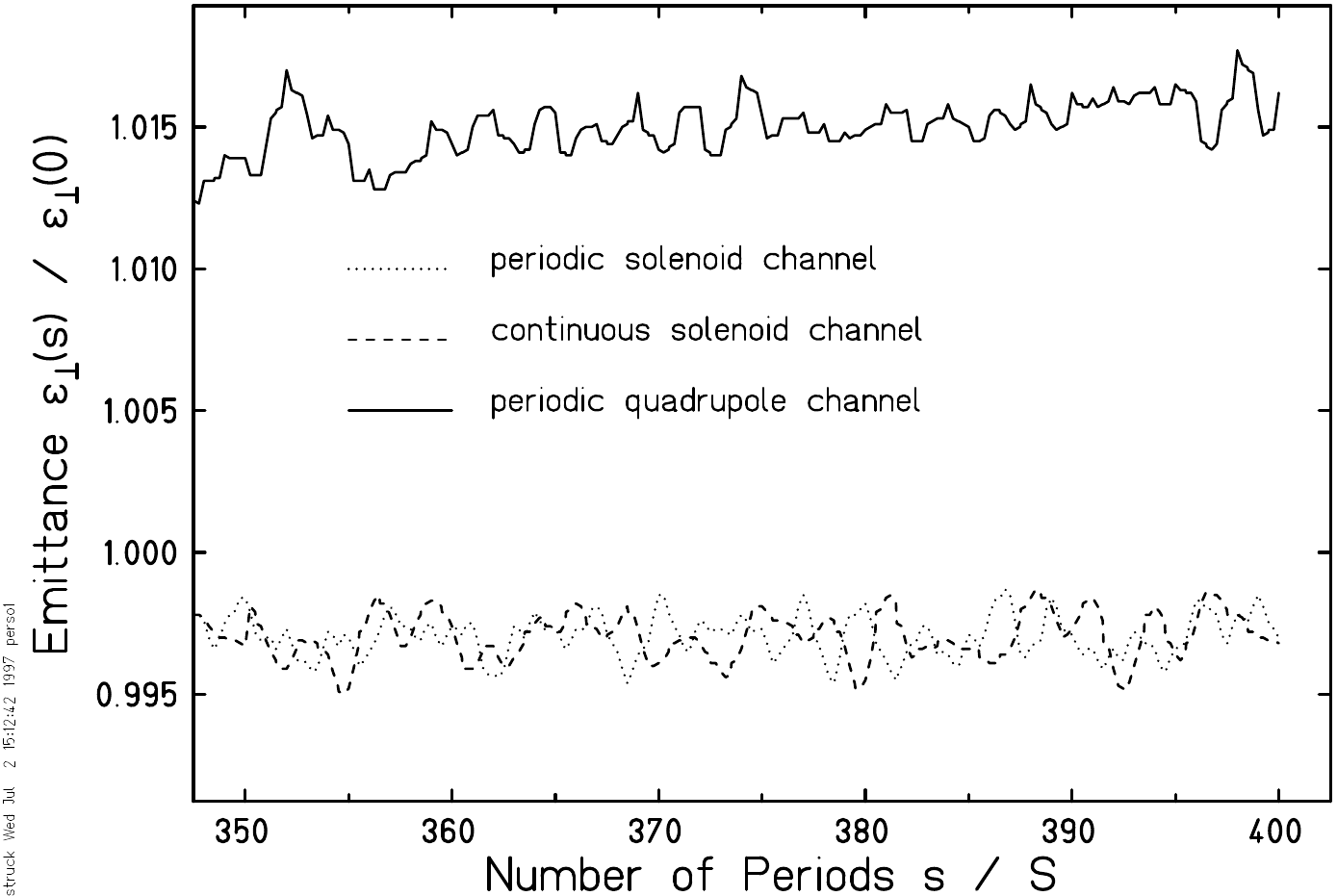,height=75mm}
\caption{Long term emittance growth factors versus the number of periods
for a stationary ``water bag'' distribution at $\sigma_0 = 60^\circ$,
$\sigma = 15^\circ$.}
\label{longterm}
\end{figure}
Fig.~\ref{longterm} shows the rms-emittance growth factors as they
occur if we transform a stationary ``water bag'' beam
over a long distance through our three types of transport channels.

Within the resolution of Fig.~\ref{longterm}, the emittance change factors
for periodic and continuous focusing solenoid channels agree closely.
The factors approach a value slightly below $1.0$, accompanied by changes
near the initially sharp boundary of the populated phase space.
If the charge density becomes less homogeneous at nearly unchanged rms size,
Eq.~(\ref{egdw}) associates the increase of excess field energy with a reduction
of the rms-emittance.
For an initially exact continuous-channel equilibrium, such changes in a
simulation require perturbations or numerical diffusion.
No further systematic emittance slope is resolved over the remaining
propagation interval displayed.

This is not true for the periodic quadrupole channel.
We observe a small but steady increase of the rms-emittance growth
factors which in our simulation example
($\sigma_0 = 60^\circ$, $\sigma = 15^\circ$) accumulates
to a factor of $1.015$ after $400$ focusing periods.

The transverse Hamiltonian generally varies in both periodic solenoid
and periodic quadrupole channels, as follows from its explicit dependence
on the focusing functions.
This variation alone does not imply secular emittance growth.
The persistent slope in the quadrupole example may reflect departures
from a periodic kinetic equilibrium, numerical effects, or both.
Its physical origin and its behavior beyond the simulated interval
require a separate convergence and stability analysis.
\section{CONCLUSIONS}
We have compared the transport of intense ion beams passing
through both continuous and periodic focusing channels.
It is already known that the motion of the beam particles in both types
of transport systems can be correlated if the equations of motion are linear.
This occurs if the particle phase space distribution is of the \kv type.
Explicitly, the relationship is given by a canonical transformation
that maps a time-dependent harmonic oscillator onto a time-independent one.
For distributions with nonlinear space-charge forces, we have identified
the limitation of extending this correspondence to a physical periodic channel.
The canonical scaling nevertheless provides useful initial conditions
whose second moments agree with those of a matched K-V reference beam.

For a common transverse scale, this matching transformation preserves
the rms-emittances and a vanishing initial rms-emittance derivative.
It does not establish exact subsequent equilibrium or exclude higher-order
emittance changes for a non-K-V distribution.
Moreover, the physical transverse Hamiltonian is generally not conserved
in a periodic focusing channel, even if the beam stays rotationally symmetric.

Our computer simulations indicate that appropriately initialized
water-bag distributions can be transported with very small rms-emittance
changes through the continuous and periodic solenoid channels investigated.
After $400$ focusing periods, the displayed rms-emittance factors in these
two channels agree within the numerical resolution.
This supports the practical usefulness of the matching prescription
for the stated parameters, without proving exact reproduction of the
full distribution or loss-free transport over arbitrary distances.

For quadrupole channels, the beam no longer stays round and its self-field
changes as the transverse aspect ratio varies.
Computer simulations show a small persistent increase of the rms-emittance
over the interval investigated.
The periodic quadrupole channel nevertheless preserves the rms properties
of the matched distribution to a high degree in these examples.
Distinguishing a physical emittance-growth mechanism from numerical
diffusion requires convergence checks and a separate stability analysis.
The reported differences therefore should not be interpreted as a general
theorem separating exact solenoid transport from inevitably degrading
quadrupole transport.

\end{document}